\documentclass[structabstract]{aa}  
\usepackage{natbib}
\bibpunct{(}{)}{;}{a}{}{,}%to folow the A&A style
\usepackage{graphicx}
\usepackage{txfonts}
\usepackage{lscape}

\begin{document}
%\texttt{}
   \title{The chemical composition of carbon stars. The R-type stars}

   \subtitle{}

   \author{O. Zamora
          \inst{1},
	  C. Abia \inst{1},
          B. Plez \inst{2}
          \and
          I. Dom{\'{i}}nguez\inst{1}
	   \and
	  S. Cristallo\inst{1,3}
          }

   \institute{Departamento de F{\'{i}}sica Te\'orica y del Cosmos, Universidad de Granada,
              18071 Granada, Spain\\
              \email{zamora@ugr.es}
         \and
             GRAAL, Universit\'e Montpellier II, CNRS, 34095 Montpellier Cedex 5, France\\
          \and{INAF, Osservatorio Astronomico di Collurania, 64100 Teramo, Italy}\\
             }

   \date{}

% \abstract{}{}{}{}{} 
% 5 {} token are mandatory
 
\abstract
%    {Late and early-R stars are giant carbon stars whose origin is not well understood.  In particular, early-R stars
% have been recently placed in the He-core burning region on the HR diagram and, thus, they are too faint to be in the AGB phase where
% carbon is expected to be mixed in the envelope by the third dredge-up.}
{}
{The aim of this work is to shed some light on the problem of the formation of carbon stars of R-type from a detailed study
of their chemical composition.}
{We use high-resolution and high signal-to-noise optical spectra of 23 R-type stars (both early- and late-types)
selected from the Hipparcos catalogue. The chemical analysis is made using spectral synthesis in LTE and 
state-of-the-art  carbon-rich spherical model atmospheres. We derive their CNO content (including the $^{12}$C/$^{13}$C ratio),
average metallicity, lithium, and light (Sr, Y, Zr) and heavy (Ba, La, Nd, Sm) $s$-element abundances. The observed
properties of the stars (galactic distribution, kinematics, binarity, photometry and luminosity) are also discussed.}
{Our analysis shows that late-R stars are carbon stars with identical chemical and observational characteristics than 
the normal (N-type) AGB carbon stars. In fact, the $s$-element abundance pattern derived can be reproduced by low-mass AGB nucleosynthesis models where the $^{13}$C($\alpha$, n)$^{16}$O reaction is the main neutron donor. We confirm
the results of the sole previous abundance analysis of early-R stars by Dominy (1984), namely: they are carbon stars with near solar metallicity showing enhanced nitrogen, low $^{12}$C/$^{13}$C ratios and no $s$-element enhancements. In addition, we have found that early-R stars have Li abundances larger than expected for post RGB tip giants. We also find that a significant number ($\sim40 \%$) of the early-R stars in our sample are wrongly classified, being probably classical CH stars and normal K giants.}
{On the basis of the chemical analysis, we confirm the previous suggestion that late-R stars are just 
misclassified N-type carbon stars in the AGB phase of evolution. Their photometric, kinematic, 
variability and luminosity properties are also compatible with this. In consequence, we suggest that the number of true R stars is considerably lower than previously believed. This alleviates the problem of considering  R stars as a frequent stage in the evolution of low-mass stars. We briefly discuss the different scenarios proposed for the formation of early-R stars. The mixing of carbon during an anomalous He-flash is favoured, although  no physical mechanism able to trigger that mixing  has been found yet. The origin of these stars still remains a mystery.}


   \maketitle
%
%________________________________________________________________

\section{Introduction} \label{intro}
Carbon stars are easily recognisable by the presence of absorp\-tion bands
of C$_2$ and CN at near visual wavelengths and  are
chemically characterised by C/O $>1$ in their envelope. The Henry
Draper classification divided carbon stars into the N and R groups on the
basis of their spectral features. The N or normal carbon stars, show a very
strong depression in the spectrum at $\lambda < 4500$ {\AA},  while the
R stars seem to be warmer and the blue/violet region of the spectrum is usually
accessible to observations. \citet{shane1928a} split the R class into R0 to R8, where
R0-4 (hot-early-R stars) are warmer and equivalent to the K-type giants, and
R5-8 (cool-late-R stars) are equivalent to M stars. Since C/O $<1$ almost everywhere
in the Universe, the carbon excess must be a result of stellar nucleosynthesis
within the star itself or in a binary companion. The N stars are understood in the
former scenario: they are luminous, cool stars with alternate H- and He-burning 
shells, that owe their carbon enhancement to the mi\-xing triggered by the third
dredge-up during the asymptotic giant branch (AGB) phase in the evolution of low-mass 
($< 3$ M$_\odot$) stars \citep[e.g.][]{iben1983}. Other carbon stars like the
classical CH-type are observed to be members of binary systems and their chemical
peculiarities can be explained by the second scenario as a consequence of mass transfer \citep{han1995}.

The R stars seem to be very common among the giant carbon stars, accounting 10 times
more than the N stars \citep{blanco1965}. According to the general catalogue of carbon
stars \citep{stephenson1973}, R-type stars may amount to $\sim 30 ~\%$ of all
the carbon stars. \citet{bergeat2002a} suggested that this fraction might be
even larger. This figure is very important since they may represent a stage of evolution that is
availa\-ble to an appreciable fraction of stars, and are not the result of anomalous initial
conditions or statistically unlikely events. Further, their velocity dispersion and position
in the Galaxy indicates that early-R stars are members of the galactic thick disk while late-R
show kinematic properties rather similar to the thin disk ste\-llar population 
\citep[e.g.][]{sanford1944,eggen1972,bergeat2002b}. The luminosities 
of R stars (at least the early-R) are known to be too low to be shell helium burning stars \citep[e.g.][]{scalo1976}. 
In this sense, a fundamental step was made by \citet{knapp2001} whose re-processing of the Hipparcos data \citep{hipparcos1997}, located
R stars in the H-R diagram at the same place than the red clump giants and concluded that they
were He-core burning stars or post helium core stars. Another fundamental property of these
stars is that no R star has been found so far in a binary system \citep{mcclure1997}, a 
statistically unlikely result. 

The only chemical analysis 
by \citet{dominy1984} showed that early-R stars have near solar metallicity,  low ($<10$) $^{12}$C/$^{13}$C  ratios, moderate
nitrogen excess and no $s$-element enhancements. No chemical analysis of late-R stars exists 
up to date. This finding for early-R stars is also in sharp contrast with that found in N-type 
carbon stars \citep{abia2002}. So, the problem is how a {\it single} giant star not luminous enough to be on the AGB phase
can present a C/O $>1$ at the surface. The favoured hypothesis so far is that the carbon produced during
the He-flash is mixed in some way to the surface. However, standard one-dimensional He-flash models 
do not predict the mi\-xing of carbon-rich material from the core to the stellar envelope \citep{harm1966}.
A lively discussion about this subject was held following the first suggestion 
of mixing at the He-flash proposed by \citet{schwar1962}. 
 Models in which rotation was parametrised \citep{mengel1976} lead 
 to off-center He-ignition but not to conditions favouring mixing. On the other 
  hand, models 
 in which the location of the He-ignition was ad hoc moved to the outer part 
 of the He-core \citep{pacinski1977} produced the desired mixing. 
  The general conclusion being \citep{despain1982} that hydrostatic models 
  do not produce mixing at the He-flash. The situation is different 
  for Z=0 models \citep{fujimoto1990,hollowell1990} but in this 
  case the mixing is directly linked to the lack of CNO elements in the H-shell
   and it can not be extrapolated to more metal-rich models\footnote{We remind that from the analysis by \citet{dominy1984},  R stars have 
[Fe/H]$\sim 0.0$. In this paper we adopt the usual notation
[X/H]$=$log N(X)/N(H)$_\star-$log N(X)/N(H)$_\odot$, where log N(H)$\equiv 12$ is the hydrogen 
abundance by number.}.
 Recently, two- and three-dimensional hydrodynamic simulations of the core He-flash in Population 
 I stars have been performed \citep{lattanzio2006,mocak2008a,mocak2008b,mocak2009}. 
None of these simulations lead to 
  substantial differences with respect 
  to the hydrostatic case. 
Thus, no physical mechanism able to
trigger the required mixing of carbon  during the He-flash in single stars has been found up to date. 
Binary star mergers have been invoked to induce such mixing. This, by passing, might explain why no R star is found to be binary.
\citet{izzard2007} investigate statistically po\-ssible channels for early-R star formation by a binary merger process.
They found many possible evolutionary channels, the most common of which is the merger of a He white dwarf with
a hydrogen-burning red giant branch star during a common envelope phase.
 However, it is far from clear if such a
merger might lead to the mixing of carbon
to the surface during the He-flash \citep{zamora2009}.

In order to shed light on the problem of the origin of these stars, we present here a detailed
chemical analysis of late- and early-R stars using high quality spectra. This is a further step in 
the study of the chemical composition of field giant carbon stars in the Galaxy. In Paper I
\citep{abia2000} we studied J-type stars and in Papers II and III \citep{abia2001,abia2002}
the N-type stars. In the next
section we describe the main characteristics of the sample stars and the observations.
Sect. \ref{estimation} describes the determination of the stellar parameters
and chemical analysis, and Sect. \ref{results} discusses the results and draws some conclusions.

%__________________________________________________________________

\section{The data}
\subsection{Observations}
We selected 23 galactic R-type stars from the sample
of carbon stars compiled by \citet{knapp2001} with  
measured para\-llaxes according to Hipparcos \citep{hipparcos1997}. 
The stars were observed with the 2.2 m telescope at CAHA observatory 
(Spain) during March 2003 and July/August 2004. Spectra were 
obtained using the FOCES echelle spectrograph \citep{pfeiffer1998}. 
The spectra cover the wavelength region from $\lambda \lambda\sim$ 4000 -- 10\,700 \AA~
in 90 orders with full spectral coverage, a resolving power of R$\sim40\,000$ in most of
the spectra except those obtained in the last period of observation (R$\sim 20\,000$), because
of the $2\times 2$ CCD binning made to increase the signal-to-noise ratio in the spectrum
of the fainter stars. The typical S/N ratio reached in the spectra was larger than 100 in the red orders and
lower than this value below $\lambda \sim 5000$ {\AA}.
In fact, in some stars the blue orders were not useful for the abundance analysis. 

\setlength\tabcolsep{0.1cm}
\begin{table*}
\begin{center}
\caption{Log of the observations and spectral classification.}
\label{obs}
\normalsize{
\begin{tabular}{c c c c c c c  l}
\noalign{\smallskip}
\hline\hline
\noalign{\smallskip}
\multicolumn{4}{c}{Name} & Var. Type&\multicolumn{1}{c}{Date} & 
\multicolumn{1}{c}{Exp.} &\multicolumn{1}{c}{Spectral Type} \\ 
HIP&BD&HD&GCVS&&(*)&(s)\\
\noalign{\smallskip}
\hline
\noalign{\smallskip}
35\,810  &$-$03\degr 1873&57\,884&V758 Mon&Lb&1&3000&N (9), R8 (12)\\
36\,623  &+24\degr 1686&59\,643&NQ Gem    &SR&1&3600&R9 (12), R6 (9), R8 (15)\\
39\,118  &-&-&-                           &-&1&3600&  R2 (15)\\
44\,812  &-&78\,278&-                     &-&1&3600&  R6 (9), R5 (15)\\
53\,832  &+41\degr 2150&-&-               &-&1&3600& R0 (15), CH-like (16), CH (2)\\
58\,786  &+71\degr 600&-&-                &-&1&3600& R2 (15), CH-like (17), CH (2)\\
62\,401  &+04\degr 2651a&111\,166&RU Vir  &Mira&1&3600& R3 (9), R3 (15)\\
62\,944  &-&112\,127&-                    &-&1&2400&K0 (13), K1 (14), K2 (11, 8, 10), K3 (7), R3 (1) \\
69\,089  &-&123\,821&-                    &-&1&3600& G9 (7), G8 (11),  R2 (4), R2 (1)\\
74\,826  &+30\degr 2637&-&-               &-&1&2700& R0 (12)\\
82\,184  &+23\degr 2998&-&-               &-&3&3600& R0 (15), R2 (1)\\
84\,266  &+42\degr 2811&156\,074&-        &-&2&2400&R0 (11, 17), R1 (5), R2 (1)\\
85\,750  &+02\degr 3336&-&-               &-&3&3600& R2 (15), CH-like (17), N5 (4), N4 (1) \\
86\,927  &+17\degr 3325&-&-               &-&2&3000&R0 (12)\\
87\,603  & -&163\,838&-&              &3&5400&R3 (9), R5 (15)\\
88\,887  &+09\degr 3576&166\,097&-        &-&2&3600& R4 (9), R5 (15) \\
91\,929  &$-$13\degr 5083&173\,138&RV Sct &Lb&3&3600& R3 (9), R3 (15), R4 (1)\\
94\,049  &$-$17\degr 5492&178\,316&-      &-&3&5400& R4 (12), R2 (9)\\
95\,422  &-&-&-                           &-&3&5400& R5 (15)\\
98\,223  &$-$00\degr 3883&188\,934&-      &-  &3&7200&R8 (12), R4 (9)\\
108\,205 &+49\degr 3673&208\,512&LW Cyg   &Lb&3&4200&R3 (9), R2 (15), R8 (3)\\
109\,158 &-&-&CT Lac&                     SRa&3&5400& R (6), N8 (3) \\
113\,150 &-&216\,649&-&                   -&3&3300&R3 (9), R5 (15)\\
\noalign{\smallskip}                                                        
\hline                                                                      
\end{tabular}
}
\end{center}
\rm
\scriptsize{* (1) 2003 March 12-13; (2) 2003 August 9-10; (3) 2004 July 1-4.\\
References for the spectral types:\\
1: \citet{barnbaum1996}; 2: \citet{bart1996};
3: \citet{eglitis2003};
4: \citet{keenan1993}; 5: \citet{keenan1941};
6: \citet{lee1944}; 7: \citet{mcclure1970};
8: \citet{morgan1973};
9: \citet{sanford1944}; 10: \citet{schild1973}; 11: \citet{schmitt1971};
12: \citet{shane1928a}; 13: \citet{stock1956}; 14: \citet{upgren1962};
15: \citet{vandervort1958}; 16: \citet{yamashita1972};
17: \citet{yamashita1975}.\\
The following stars show evidence of binarity:\\ HIP 36\,623 \citep{carquillat2008}, HIP 109\,158 \citep{makarov2005}, HIP 53\,832 \citep{platais2003} and
HIP 85\,750 \citep{mcclure1997}.}
\end{table*}

The spectra were reduced using the {\it echelle} task within the IRAF package following the standard procedure.
When several images of the same object were obtained, they were reduced independently and finally added
using the IRAF task {\it scombine}. The wavelength calibration was always better than 0.03 \AA. Finally,
the individual spectral orders were normalised to a pseudo-continuum by joining the maximum flux points and, in some
cases, corrected with the help of the synthetic spectra (see details of the method in Paper I). 
We estimate an uncertainty in the continuum location of less than a $5\%$, although it may be larger for the
spectral orders severely affected  by mole\-cular bands, particularly in late-R stars.

\subsection{Spectral classification}

One of the most serious problems when dealing with carbon stars in general is their
ascription to one or another spectral type. This is because the spectral classification is done
frequently by using low resolution spectra, which does not permit the detection of
relevant atomic lines due to blends with molecular bands 
\citep[e.g.][]{cannon1918,shane1928a,keenan1941,vandervort1958,keenan1993,barnbaum1996}. In this sense, 
studies on early-R stars are frequently contaminated with CH-type stars, whose spectral distribution is
very similar \citep[see e.g.][]{wallerstein1998}. In fact, the most 
effective way of distinguishing between these two spectral types is to compare the intensity of some metallic lines (less prominent
in CH-type stars) and, mainly, the intensity of $s$-element lines which are well known to be 
enhanced in CH-type stars \citep[][]{keenan1993,abia2003,goswami2005}. An additional  problem concerns the derivation 
of the temperature subtype. In general the intensity of C-bearing molecular bands in carbon stars depends not only on 
the effective temperature but also on the actual C/O ratio in the atmosphere. In the case of early-R stars this introduces an 
extra uncertainty in the assignation of the temperature subtype compared with normal (O-rich) G-\index{} and K-type stars of similar effective temperature. For the late-R stars, there is no tight correlation between the spectral distribution and effective temperature due to the presence of very intense molecular absorptions, similar to N-type stars \citep{keenan1993}. Table \ref{obs} shows
the spectral classification of our stars from different sources in the literature. It is evident that many stars have been classified in very different ways because of the problems cited above.

This confusion in the spectral classification of many R-type stars has motivated the use of others criteria such as the
photometric one. For instance, \citet{knapp2001} use the $(V-K)$ colour index to distinguish between early 
and late-type R stars according whether $(V-K)_0$ $<$ 4 or $(V-K)_0$ $>$ 4, respectively. Since many colour indexes 
might be affected by the variability of the star (at least for the late-R) and also by the
presence of strong molecular absorptions, we simply adopt here the criterion based 
on the effec\-tive temperature estimated in the chemical analysis (see Sect. \ref{estimation}). We will see, that an effective temperature thres\-hold of $\sim 3600$ K seems to be a good criterion to distinguish between early- and late-R stars. This will 
lead to a spectral classi\-fication in agreement with the \citet{knapp2001}'s criterion above, and with other
characteristics that differentiate the two subtypes of R stars.

\subsection{Distribution in the Galaxy and kinematics} \label{distri}

That the R stars are galactic disk objects was recognised by \citet{eggen1972}. \citet{bergeat2002b} calculated the space density
in the galactic plane of early-R stars and found that it is a factor $\sim 16$ lower than for N stars; in fact, on average they
are three times further from the galactic plane. Despite the small sample studied here, we reach the same conclusion: our early-R
stars are located at large galactic latitudes, $\mid$b$\mid$ $\geq$ 30$\degr$. The distribution of the late-R stars 
is however, almost identical to that of the N-type stars \citep[e.g.][]{claussen1987}: they are very close to the galactic plane. 
These differences in the galactic distribution of R stars were already known \citep[e.g.][]{sanford1944,ishida1960,rybski1972,stephenson1973,barbaro1974} and are indicative of 
early-R stars belonging to the galactic thick disk while late-R stars belong to the thin disk. This of course implies a range for the masses and ages of the R stars: early-R stars must be of lower masses ($\leq 1$ M$_\odot$) and older than late-R stars. 

Previous kinematics analyses performed by e.g. \citet{dean1972} and recently by \citet{bergeat2002b} show that the velocity dispersion 
of R-stars are typically larger by a factor of $\sim 2$ than for N-type stars.
\citet{bergeat2002b} in particular, obtain $\sigma = 42 - 54$ km s$^{-1}$ in the direction of the galactic
north pole for the stars belonging to their {\it hot carbon} group\footnote{Bergeat et al. (2002b) define 
14 photometric groups in order to classify the carbon stars observed by Hipparcos into homogeneous classes. 
They define the \textit{hot carbon} (HC) group which includes mostly early-R and CH-type stars, whereas
the \textit{cool variables} (CV) group includes N-type stars and a few late-R stars.}, where most of our 
early-R are included,  and 23 km s$^{-1}$ typically for the {\it cool variable} group, which includes the late-type stars 
in our sample. This reinforce the conclusion that the differences between early- and late-type R stars are
representative of two different stellar populations, as we already noted. According to this numbers, 
and applying a standard age-velocity relation \citep[e.g.][]{wielen1992}, an age of $\sim 3$ Gyr and $\gtrsim 10$ Gyr is
obtained for late- and early-R stars, respectively.

\setlength\tabcolsep{0.12cm}
\begin{table}
\begin{center}
\caption{Photometric data and luminosities}
\label{magnitudes}
\normalsize{
\begin{tabular}{c c c c c  c c c   c c}
\noalign{\smallskip}
\hline\hline
\noalign{\smallskip}
Star &$V$&$K$&$J-H$&$H-K$&$M_{K_0}$&$M\rm{_{bol}}$&Ref.\\
\hline
\noalign{\smallskip}
\multicolumn{8}{c}{Late-R stars} \\
\hline
\noalign{\smallskip}  
HIP 35\,810  &  9.01& 3.67& 0.83& 0.42 & $-$6.84& $-$4.05 &   1        \\                                            
HIP 36\,623  &  8.02& 2.95& 0.72& 0.38 & $-$7.23& $-$4.18 &   5        \\                                            
HIP 62\,401  & 11.98& 1.81& 1.85& 1.30 & $-$7.48& $-$5.23 & 7         \\                                            
HIP 91\,929  &  9.75& 3.41& 1.30& 0.51 & $-$7.25& $-$4.96&    1       \\                                            
HIP 108\,205 &  9.23& 1.71& 1.36& 0.80 & $-$8.57& $-$5.45&   4, 1        \\                                            
HIP 109\,158 & 10.12& 2.57& 1.21& 0.75 & $-$8.90& $-$5.65&    5, 1       \\                                           
\hline\noalign{\smallskip}
\multicolumn{8}{c}{Early-R stars} \\
\hline
\noalign{\smallskip}
HIP 39\,118*  & 10.41&  7.23& 0.79& 0.25 &  $-$2.18& 0.47      &  1        \\                                            
HIP 44\,812  & 10.61 &  6.50& 0.78& 0.35 &  $-$4.73 &$-$2.02  &  1          \\                                            
HIP 53\,832  & 10.11 &  7.60& 0.52& 0.13 &  $-$2.83 &$-$0.81  &  1          \\                                            
HIP 58\,786  & 10.27 &  7.51& 0.52& 0.15 &  $-$3.06 &$-$1.02  &  1         \\                                            
HIP 62\,944*  &  6.91&  4.16& 0.55& 0.10 &  $-$1.27& 0.86      &  3          \\                                            
HIP 69\,089*  &  8.68&  6.40& 0.46& 0.12 &$-$3.09   & $-$1.24        &  1         \\                                            
HIP 74\,826  &  9.78 &  7.39& 0.48& 0.12 &  $-$3.31 & $-$1.38 & 1          \\                                            
HIP 82\,184  &  9.10 &  6.46& 0.57& 0.12 &  $-$3.67 & $-$1.04 & 1           \\                                            
HIP 84\,266  &  7.60 &  5.11& 0.30& 0.23 &$-$3.09    & $-$0.87 & 5         \\                                            
HIP 85\,750  &  9.42 &  5.15& 0.82& 0.24 &$-$5.85    & $-$3.25 & 2, 1       \\                                            
HIP 86\,927  &  8.71 &  6.14& 0.53& 0.14&  $-$1.69    & 0.22    & 2           \\                                            
HIP 87\,603  & 10.72 &  7.92& 0.55& 0.18&  $-$2.24    & $-$0.02 & 1         \\                                            
HIP 88\,887  &  9.78 &  5.13& 0.84& 0.36&  $-$5.90    & $-$3.14 & 1       \\                                            
HIP 94\,049  & 10.39 &  7.39& 0.55& 0.19  &$-$3.82   & $-$1.97 & 1         \\                                            
HIP 95\,422  & 11.01 &  6.74& 0.75& 0.26 &  $-$4.81   & $-$2.57 & 1          \\                                            
HIP 98\,223  &  9.36 &  5.15& 0.78& 0.31  &  $-$6.27  &$-$3.78  &6, 1        \\                                            
HIP 113\,150 & 10.75 &  7.97& 0.55& 0.22 & $-$3.45   &$-$1.26  & 1             \\             
\hline
\noalign{\smallskip}
\end{tabular}
\rm
}
\end{center}
\scriptsize{Non dereddened $VJHK$ magnitudes. \\
References of photometry:\\
1: 2MASS \citet{2mass}; 2: \citet{dominy1986}; 3: \citet{elias1978}; 4: \citet{neugebauer1969};
5: \citet{noguchi1981}; 6: \citet{mendoza1965}; 7: \citet{whitelock2000}.
}
\end{table}

\subsection{Binarity}

Table \ref{obs} indicates the stars with evidence of binarity obtained from radial velocity
variations or other methods. These stars are HIP 53\,832 and HIP 85\,750,  classified as early-R, and the 
late-R stars HIP 36\,623 and HIP 109\,158. It is important to note that these two early-R stars have been also classified 
as CH-type stars, which are known to be all binaries. We will reach the same conclusion on the basis of their
chemical composition (see Sect. 4.1). HIP 85\,750 is a binary system whose orbital parameters were determined by \citet{mcclure1997}.
The late-R star HIP 36\,623 is a symbiotic star detected by time-variations of the ultra-violet 
continuum due probably to the presence of a white dwarf companion 
\citep{johnson1988,belczynski2000,munari2002}. Recently, \citet{carquillat2008} estimate that the 
components of the system have masses of 2.5 M$_{\sun}$ and 0.6 M$_{\sun}$. On the other hand, HIP 109\,158 appears in the
catalogue of Hipparcos astrometric binaries with accelerated proper motions \citep{makarov2005}. For the remaining sample, 
there is no evidence of binarity (in agreement with the study by \citealt{mcclure1997}) or available study.

\subsection{Photometry} \label{photometry}

\begin{figure}
\begin{center}
\includegraphics[width=9cm,angle=0]{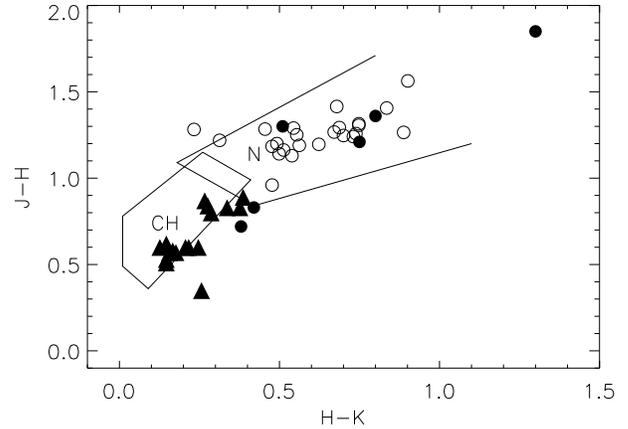}
\protect\caption{Near infrared colour-colour diagram for the stars in this study. Triangles are early-R stars, filled 
circles late-R stars and empty circles the N-type stars from \citet{abia2001}. The regions typically 
populated by CH- and N-type stars are indicated \citep{totten2000}.}
\label{fig2mass}
\end{center}
\end{figure}

Table \ref{magnitudes} shows the available  $VJHK$ photometry taken from \citet{knapp2001}, the Hipparcos catalogue or
from the 2MASS ({\it Two Micron All Sky Survey}) on-line data release \citep{2mass}. Differences in the photometry found in the literature for a given
star are typically below $\sim$ 0.2 mag for the majority of early-R stars, but it can be higher for late-R stars due to their known pho\-tometric variability (see Table \ref{obs}). All the pho\-tometric magnitudes were corrected from 
interstellar extinction using the $A_J$ values derived by \citet{bergeat2002a} and the relations by \citet{cardelli1989}.
For the stars HIP 39\,118, HIP 62\,944 and HIP 69\,089, not included in the Bergeat et al. (2002a) sample, we use
the galactic interstellar extinction model by \citet{arenou1992} and the parallaxes by Knapp et al. (2001).
In Fig. \ref{fig2mass} the position of our sample stars is plotted in the near infrared colour-colour diagram. We have marked with lines 
the regions usually occupied by CH- and N-type stars according to \citet{totten2000}. Some galactic carbon stars of N-type 
(Paper II) are included for comparison. It is clear from this fi\-gure that early-R stars occupy the same region
as CH-type stars, while late-R stars are redder and are located mostly in the region of N-type carbon stars. 

Another interesting photometric index is the $K-[12]$\footnote{ $[12] =- 2.5$ log $(F12/28.3)$, where $F12$
is the IRAS flux density at 12 $\mu$m.} 
colour, which is considered a tracer of the dust surrounding the star \citep{jorissen1998}. The formation of circumstellar dust is considered a measure of the stellar mass-loss and in consequence might give an indication of the
evolutionary status of a given star: stars with no mass-loss or very low rates have $K-[12] \leq 0.7$.
Our early-R stars fulfil this criterion and do not present significant mass-loss.
At contrary, most of our late-R stars show $K-[12] > 0.7$ and would correspond to stars with moderate mass-loss 
rates, \mbox{5$\times$10$^{-8}$} to \mbox{5$\times$10$^{-7}$} M$_{\sun}$ y$^{-1}$, the extreme case being HIP 62\,401 ($K-[12]= 4.1)$  
having a considerably larger rate, 10$^{-6}$ M$_{\sun}$ y$^{-1}$ (the star in the upper right corner in Fig. \ref{fig2mass}). These numbers are similar to those estimated in normal AGB stars 
\citep[see e.g.][]{busso2007a}.

\subsection{Luminosity} \label{luminosity}

As we mentioned above, the selected stars have all measured trigonometric parallaxes according to
Hipparcos \citep{hipparcos1997}. However, Hipparcos parallaxes for giant stars resulted to be not very accurate. 
\citet{knapp2001} reprocessed the original Hipparcos parallaxes by imposing the condition that they remained always greater than zero 
\citep[see also][]{pourbaix2000}. In the sample of R stars in \citet{knapp2001} only 18 $\%$ of the stars have
$\pi/\epsilon(\pi)$ $>$ 2, where $\pi$ denotes the trigonometric parallax and $\epsilon(\pi)$ the associated error.
\citet{knapp2001} performed a re-reduction of the \textit{Intermediate Astrometric Data} \citep[IAD,][]{vanleeuwen1998}
taking as parameter, in the $\chi^2$ minimisation, the logarithm of the distance instead of the distance itself. In this way, the derived parallaxes are always greater than zero. The true parallaxes, i.e. the parallaxes free of all the possible biases \citep[see e.g.][]{arenou1999}, were then estimated with a Monte-Carlo simulation by adopting di\-fferent distributions for the abscissa residuals of the IAD \citep[see also][]{pourbaix2000}. Following this procedure, these authors derived a true 
average absolute $K$-magnitude for the early-R stars: ${M_{K_O}}$ = $-2 \pm 1$, similar to the
value $-1.61 \pm 0.03$ found by  \citet{alves2000} for red clump giant stars near the Sun observed by Hipparcos.
Another major study on the parallaxes of R stars was made by \citet{bergeat2002a}. These authors use the original 
Hipparcos parallaxes and correct them from all the observational biases (that may introduce an uncertainty up to
$\sim$ 0.4 mag in the estimation of the luminosity, see e.g. \citealt{bergeat2002a} for further details),
obtaining significantly different results. In fact, they find an ave\-rage $K$-absolute magnitude of $M_{K_0} = -3.0$ for
early-R stars. Correspondingly, they derive also bolometric magnitudes for R type stars $\sim 1-2$ mag brighter
than those obtained by \citet{knapp2001}. We believe that the parallaxes by \citet{bergeat2002a} are more accurate 
since the parallaxes by \citet{knapp2001} might undergo a bias when $K \gtrsim$ 8, whose effect is to overestimate 
the $M_K$ (see their Fig. 2). On the other hand, a new analysis of the Hipparcos data has been performed recently 
by \citet{vanleewen2007}.
The absolute magnitudes computed with these new data for our stars (13 of them) are in general in good agreement with 
those derived by \citet{bergeat2002a}, with a maximum difference below 0.5 magnitudes.
Therefore, we have adopted here the parallaxes derived by \citet{bergeat2002a} to estimate the luminosity of our stars.
For HIP 39\,118, HIP 62\,944 and HIP 69\,089 (not included in the \citealt{bergeat2002a} sample, as previously noted) we use instead \citet{knapp2001} parallaxes.

\begin{figure}
\begin{center}
\includegraphics[width=9cm,angle=0]{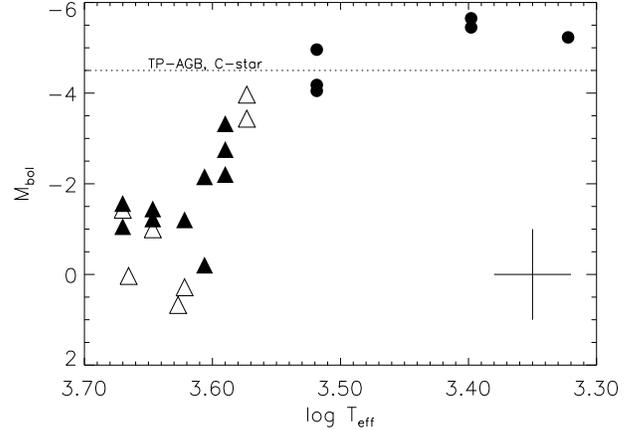}
\protect\caption{H-R diagram for the sample stars. Solid triangles are early-R stars and circles late-R stars.
Open triangles correspond to stars classified as early-R stars in the Hipparcos catalogue which have been
reclassified here as K- or CH-type giants (see Sect. 4).}
\label{hrd}
\end{center}
\end{figure}

Table \ref{magnitudes} shows the absolute magnitude in the $K$ band corrected for interstellar extinction (${M_{K_O}}$). 
The corresponding absolute bolometric magnitudes ($M\rm{_{bol}}$) according to \citet{bergeat2002a} are also shown. For the stars 
HIP 39\,118, HIP 62\,944 and HIP 69\,089 we used the bolometric correction in $K$ by \citet{costa1996}. We note that the error in 
the derivation of $M_{K_0}$ and $M\rm{_{bol}}$ can be as large as $\sim 1-1.5$ mag, and it is dominated by the uncertainty in 
the trigonometric parallaxes. Nevertheless, we have constructed the H-R diagram
of the stars in the sample (see next section for details in the derivation of the $T\rm{_{eff}}$). From
Fig. \ref{hrd} it is clear that the luminosities of late-R stars are close to the values expected for low-mass stars 
($<2$ M$_\odot$) in the AGB phase. In Fig. \ref{hrd} is also drawn the minimum luminosity (dashed line at 
$M\rm{_{bol}}\approx-4.5$) for a 1.4 M$_\odot$ star to become a thermally-pulsing AGB carbon star according to \citet{straniero2003}.
 This luminosity limit is quite model
dependent (metallicity, mass loss history, third dredge-up efficiency, treatment of the opacity in the envelope etc, see \citealt{straniero2003}), but roughly indicates that late-R stars have luminosities compatible with the AGB values. 
At contrary, early-R stars are clearly below the expected luminosity in this phase. 

\section{Atmospheric parameters and abundance calculations} \label{estimation}

\begin{figure*}
\centering
\includegraphics[width=17cm,angle=0]{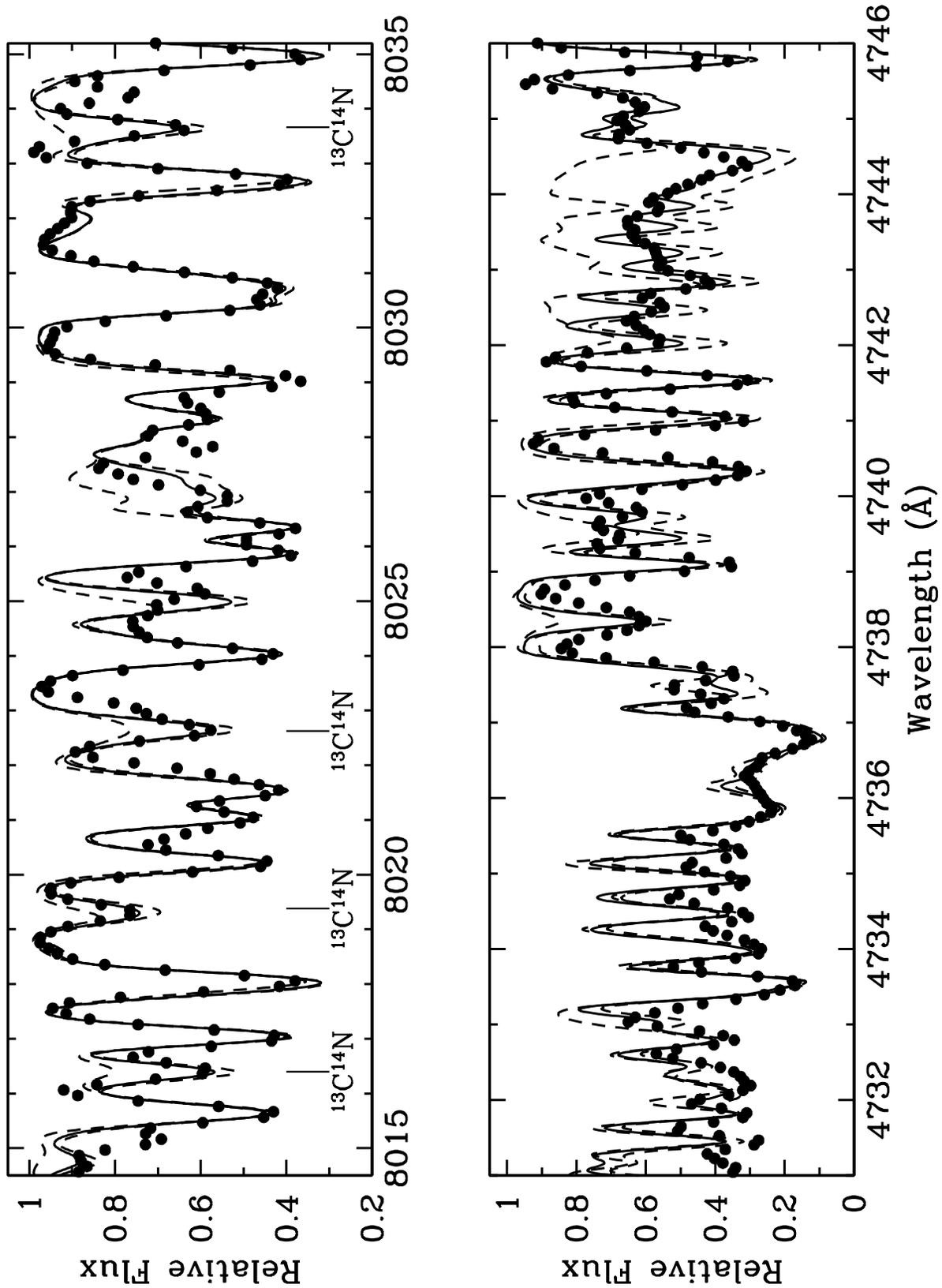}
\caption{Observed (filled circles) and synthetic spectra (continuous and dashed lines) of the early-R star HIP 84\,266 around 
the $\lambda$ 8015 {\AA} (top), and the $\lambda$ 4740 {\AA} C$_2$ band regions (below), respectively. In both spectral ranges the synthetic spectrum giving the best fit (continuous line) was computed with C/N/O = 8.78/8.40/8.56 and 
$^{12}$C/$^{13}$C$ = 7$. The position of some $^{13}$C$^{14}$N lines used to derive the carbon isotopic ratio in the 8015 {\AA} region are indicated. Dashed lines in both panels correspond to 
synthetic spectra computed with $^{12}$C/$^{13}$C = 4 and $^{12}$C/$^{13}$C = 100, respectively.}
\label{hip84266_CN0}
\end{figure*}  

We use the latest generation of MARCS spherically-symmetric model atmospheres for carbon stars \citep{gustafsson2008}. 
In this new grid however, there is no a complete set of carbon enhanced models for $T\rm{_{eff}} \geq 3500$ K. For stars with
effective temperature larger than this value, we use MARCS C-rich models computed by one of us (B. Plez).

The estimation of the atmospheric parameters was made in a similar way to
Papers II and III, and will be only briefly described here. We refer to these previous
works for more details. We followed an iterative method
that compares computed and observed spectra modifying $T\rm{_{eff}}$, gravity, average metallicity and 
the C/O ratio from given initial va\-lues until a reasonable fit to the observed spectrum is found. For the effective temperature, initial values were derived as the average temperature obtained from the calibrations 
of $(V-K)_0$, $(H-K)_0$ and $(J-K)_0$ vs. $T\rm{_{eff}}${} according to \citet{bergeat2001}. For a given
object, the typical dispersion in the $T\rm{_{eff}}$ derived from the three indexes was $\pm$200 K for the late-R stars
and $\pm$500 K for early-R stars. However, in the case of the early-R stars, the theoretical spectra are so 
sensitive to a small variation of the $T\rm{_{eff}}$, that the uncertainty in the temperature is
 certainly much lower than $\sim \pm$ 500 K.
The spectroscopic
method of deriving the effective temperature is not recommended in our case because of the blending with molecular lines, even in the warmer early-R stars. Further, the range in excitation energies of the
apparently clean atomic lines available in our spectra is too narrow. Nevertheless, we checked that for 
the warmest early-R stars where the molecular contribution can be considered weak, we did not find any
correlation between the estimated abundance of iron and the corresponding excitation
energy of a number of Fe I lines (for instance in HIP 84\,266, HIP 69\,089 and HIP 39\,118).

Gravity was estimated using the relation between lumino\-sity, effective temperature and stellar mass.
We adopted a mass of 1 M$_{\sun}$ for all the stars in the sample. Although late-R stars are probably more massive (see Sect. 2.3), an uncertainty of a factor of two in the stellar mass translates to an uncertainty of $\sim\pm 0.3$ dex in the gravity. Obviously, the most 
important source of uncertainty in the derived gravity is the luminosity. We estimate a maximum uncertainty in
log g of $\sim \pm 0.9$ dex due to the combined uncertainty in the mass, effective
temperature and luminosity. However, the theoretical spectra are very sensitive to such a variation in gravity, so the
actual uncertainty in gravity is actually lower than this maximum value ($\sim \pm 0.5$).
The mean difference between the estimated gravity and the final adopted value in the analysis was
$\sim \pm 0.1$ dex for early-R stars. For late-R stars, the differences are below 0.5 dex except in the case
of HIP 108\,205 (0.9 dex). 

We adopted initially a solar metallicity for all the stars acco\-rding to the previous
analysis by \citet{dominy1984}. The reference solar abundances are those from \citet{asplund2005}.
The final metallicity of the star ([M/H]) was obtained as the average derived from a number of
Fe, Ni, Mn and Zn lines. For solar-metallicity and moderate metal-poor stars these species scale 
approximately with Fe (i.e. [X/Fe]$\sim 0.0$), thus, these elements can be safely used as metallicity indicators.
The metallicity deduced from the [X/H] ratios (with X = Fe, Ni, Mn and Zn) agreed within $\pm$0.05 for the early-R stars and $\pm$0.10 for the late-R stars.
 A microturbulence parameter of 2 km s$^{-1}$ was adopted in all the stars, which
is a typical value for giant stars. This initial value was modified in the chemical analysis by fitting the profile
of metallic lines apparently free of molecular blends. 
We note that the microturbulence velocity giving the best fit to the observed spectrum 
may change slightly with the wavelength range. The average microturbulence velocities are given in Table \ref{adopted}. 
The macroturbulence velocity adopted was between 5 $-$ 10 km s$^{-1}$. Finally, the theoretical spectra were
convolved with a gaussian distribution to mimic the intrumental profile. The final FWHM values adopted were within $\sim$ 200 -- 400 m\AA{}, 
depending on the wavelength range and the spectral resolution.

\setlength\tabcolsep{0.1cm}
\begin{table}
\begin{center}
\caption{Atmospheric parameters and C, N, O abundances}
\label{adopted}
\small{
\begin{tabular}{c c c c c c c c}
\noalign{\smallskip}
\hline\hline
\noalign{\smallskip}                                                                                                                              
Star &$T\rm{_{eff}}$ (K)&log g&$\xi$ (km s$^{-1}$)&[M/H]&C&N&O\\
\noalign{\smallskip}
\hline\noalign{\smallskip}
\multicolumn{8}{c}{Late-R stars}\\
\hline\noalign{\smallskip}
HIP 35\,810   &3300 &0.0 &2.8&--0.38 &8.40  &7.30$^a$  &8.35        \\
HIP 36\,623   &3300 &0.0 &2.1&--0.27 &8.35 & 7.70  &8.31         \\
HIP 91\,929   &3300 &0.0 &2.5&0.00   &8.68 & 7.78$^a$ &8.66    \\
HIP 108\,205  &2500 &0.0 &2.3&0.00   &8.67 & 7.78$^a$ &8.66    \\
HIP 109\,158  &2500 &0.0 &2.5&--0.02 &8.68 & 7.78$^a$ &8.66     \\
\hline\noalign{\smallskip}
\multicolumn{8}{c}{Early-R stars}\\
\hline\noalign{\smallskip}
HIP 39\,118 &4250&2.0 &1.8&--0.29  & 8.77 & 8.00&$<$9.05$^b$  \\
HIP 44\,812 &3950&1.5 &3.0&--0.03  & 8.75 & 8.70&8.66  \\
HIP 53\,832 &4500&2.5 &2.3&--0.77  & 8.08 & 8.10&8.06  \\
HIP 58\,786 &4250&2.0 &2.2&--0.29  & 8.63 & 7.60&8.47  \\
HIP 62\,944 &4300 &2.4 &2.0&0.12   & 8.80 & 8.30&9.05$^b$  \\
HIP 69\,089 &4750 &1.5 &2.0&--0.17 & 8.40 & 8.40&8.46  \\
HIP 74\,826 &4750 &2.0 &1.5&--0.30 & 8.43 & 8.18&8.36  \\
HIP 82\,184 &4500 &2.0 &3.0&--0.15 & 8.42 & 8.25&8.51  \\
HIP 84\,266 &4750 &2.0 &2.4&--0.10 & 8.78 & 8.40&8.56  \\
HIP 85\,750 &3800 &1.2 &2.0&--0.48 & 8.38 & 7.98&8.36  \\
HIP 86\,927 &4700 &2.4 &2.3&--0.05 & 8.47 & 7.98&8.66  \\
HIP 87\,603 &4100 &2.0 &2.3&--0.52 & 8.60 & 7.68&8.36  \\
HIP 88\,887 &3950 &1.5 &3.0&--0.09 & 8.80 & 8.70&8.66  \\
HIP 94\,049 &4100 &2.0 &2.3&--0.62 & 8.48 & 7.98&8.36  \\
HIP 95\,422 &3950&2.0  &3.0&--0.26 & 8.73 & 7.78$^a$&8.36  \\
HIP 98\,223 &3800&1.5  &1.9&--0.79 & 8.05 & 7.10&7.97  \\
HIP 113\,150&4500&2.0  &2.4&--0.47 & 8.85 & 7.60$^a$&8.36  \\
\hline
\end{tabular}
}
\end{center}
\scriptsize{$^a$ N abundance scaled with the stellar metallicity. \\
$^b$ O abundance derived from the $\lambda$ 6300.3 \AA{} \mbox{[O I]} line using the atomic line parameters 
from \citet{caffau2008}.\\}
\end{table}

In the analysis of carbon stars a critical issue is the determination of the C, N, and O abundances since
the actual C/O ratio in the atmosphere determines the dominant molecular bands and thus the global
spectrum. The actual N abundance has a minor role in this sense. Unfortunately the O abundance
is very difficult to measure in carbon stars at visual wavelengths. Only in two early-R stars in our sample, HIP 39\,118 and HIP 62\,944, it was possible to derive the O
abundance from the $\lambda$ 6300.3 \AA{} $\rm{[O I]}$ line. For the other stars, the O abundance was scaled with 
the ave\-rage metallicity. We note that \citet{dominy1984} derived almost scaled to solar O abundances 
from molecular lines in the infrared in his sample of early-R stars. Nevertheless, the effect of the absolute O abundance 
on the theoretical spectrum is secondary with respect to the actual C/O ratio. This is because
for a given C/O ratio (or difference [C/H] $-$ [O/H]), there is a range of absolute O abundances (within a factor of three) 
giving almost equal theoretical spectra \citep[see][]{delaverny2006}. To estimate the C and N abundances, we proceeded as follow:
first, the carbon abundance was derived from the C$_2$ Swan band lines at $\lambda\lambda$ 4730 -- 4750 \AA{} 
taking the nitrogen and oxygen abundances scaled with the average metallicity (except in the case of O for the two early-R stars
mentioned above). The intensity of these C$_2$ lines are not very sensitive to the N and O abundances adopted. From this 
region we also derive the $^{12}$C/$^{13}$C ratio from some available $^{12}$C$^{13}$C and $^{13}$C$^{13}$C lines. 
Next, we synthesize the region at $\lambda\lambda$ 8000 -- 8050 \AA{}, which is dominated by the red system of
CN molecule, to estimate the N abundance adopting the C abundance from the previous step. Then, we returned to
the $\lambda\lambda$ 4730 -- 4750 \AA{} region to check the consistency of the theore\-tical and observed spectrum with the new N 
abundance derived. This procedure was repeated several times until convergence was found. Typically two or three
iterations were needed. For a given star, the maximum difference between the C and  N abundances derived from the 
red and blue parts of the spectrum was less than 0.2 or 0.1 dex for the early- and late-R stars, respectively.
The final carbon isotopic ratio obtained is the average value between those obtained from the fits to the 
$\lambda\sim$ 8000 \AA{} and $\lambda\sim$ 4750 \AA{} regions. The final C, N and O abundances together
with the other atmospheric parameters are shown in
Table \ref{adopted}.

\begin{figure}[t!]
\centering
\includegraphics[width=7cm,angle=-90]{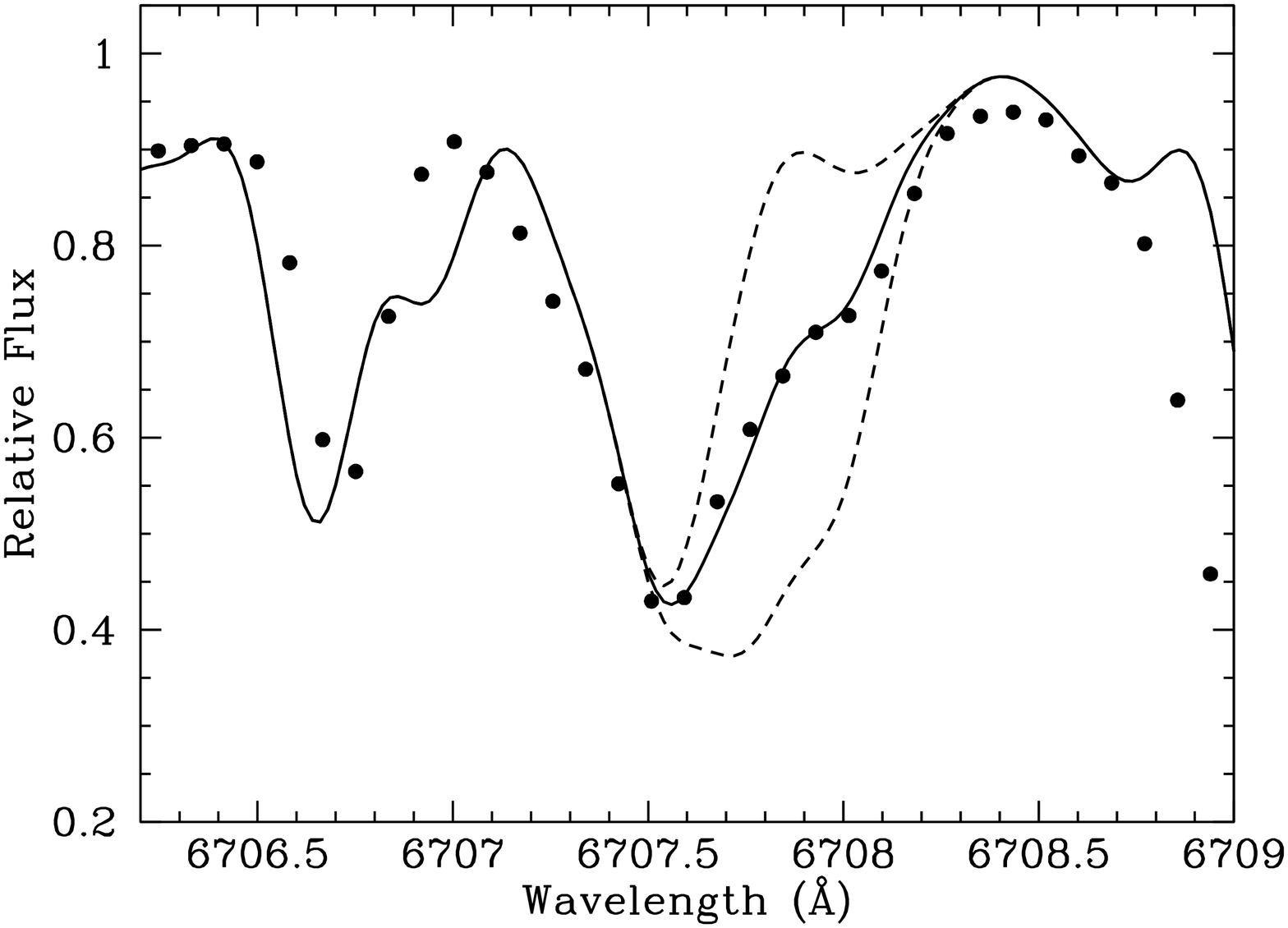}
\caption{Observed and synthetic spectra in the region of the Li line at $\lambda \sim$ 6707.8 \AA~for 
the early-R star HIP 58\,786. Symbols as in Fig. \ref{hip84266_CN0}. Continuous line is the best fit with  log $\epsilon$(Li)$ = 1.0$,
and dashed lines are theoretical spectra computed with log $\epsilon$(Li)$ = 0.0$ and 1.5, respectively.}
\label{liabundance}
\end{figure}  

\begin{figure*}[ht!]
\centering
\includegraphics[width=17cm,angle=0]{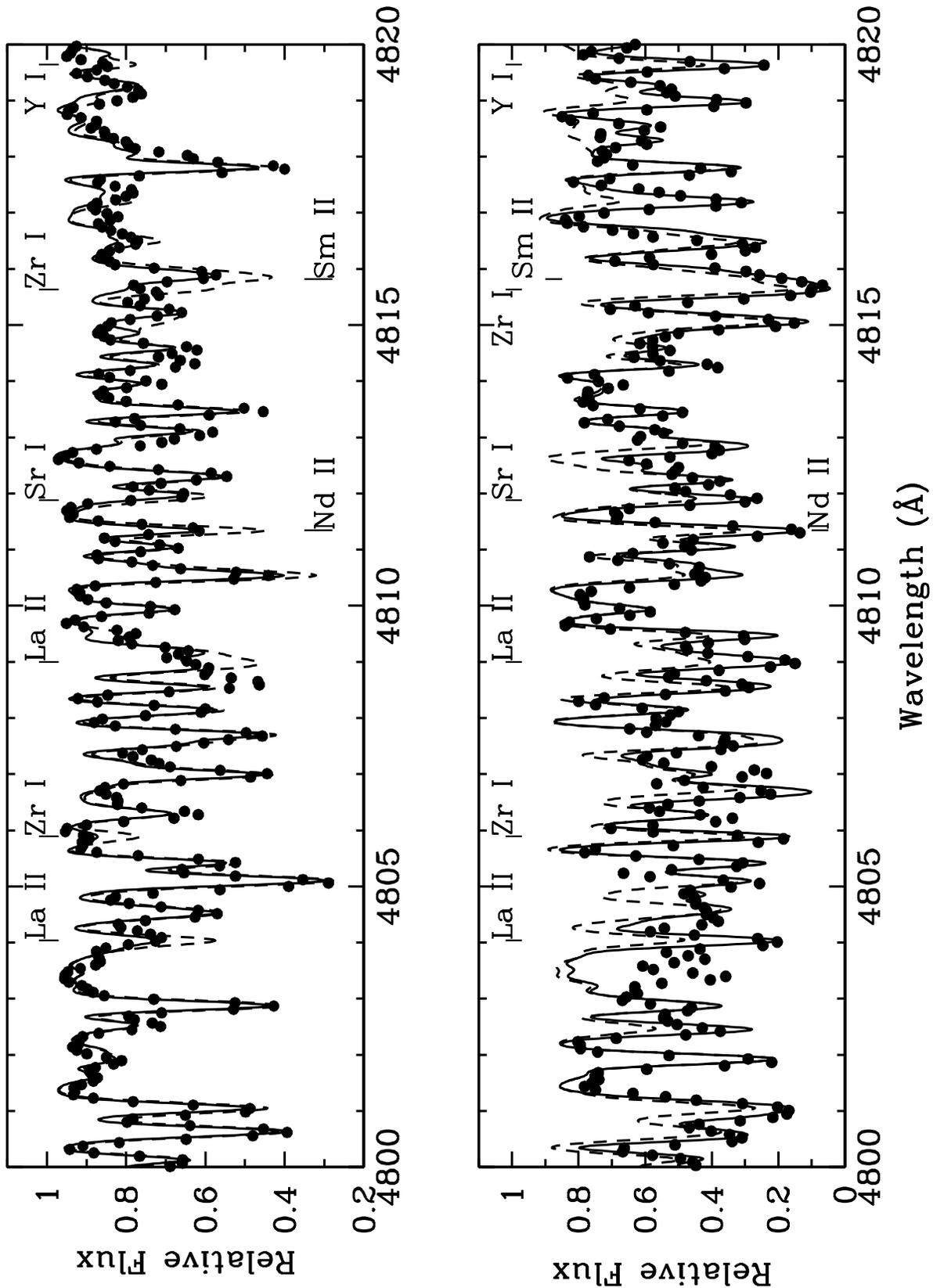}
\caption{Observed and synthetic spectra in the $\lambda$ 4800 {\AA} region for the early-R star HIP 84\,266 (top panel)
and the late-R star HIP 36\,623 (bottom). Symbols as in Fig. \ref{hip84266_CN0}. For each star two theoretical fits are shown: 
continuous line (best fit) is computed assuming [s/M] $\sim0.0$, while dashed line with [s/M] $\sim +0.5$ in the
star HIP 84\,266. For HIP 36\,623, continuous line  is computed with [s/M] $\sim +1.3$ and dashed line with [s/M] $\sim 0.0$.}
\label{late_s}
\end{figure*}  

Synthetic spectra in LTE were computed for a specific star using the 
version 7.3 of the \textit{Turbospectrum} code \citep{alvarez1998}. 
We used the same extensive set of atomic and molecular lines as
in Papers II and III and refer to these works for details. The list
of lines is available from the authors upon request.
Figures \ref{hip84266_CN0}  to  \ref{late_s}  show examples of theoretical fits in four different spectral regions used
in the chemical analysis. The lithium abundance was derived using the resonant Li I at $\lambda$ 6707.8 \AA. This 
spectral range is also interesting because of the presence of one useful Y I line at $\lambda~6687.5$ \AA{}. For most of the
stars in the sample, the Li abundance has 
to be considered only as an upper limit. This is due to the severe blending in this spectral region, as can be
seen clearly in Fig. \ref{liabundance}. $s$-element abundances were determined mainly using the spectral 
window at $\lambda \lambda$ 4750 -- 4950 \AA{} (see Papers II and III) where several lines of Sr, Y, Zr, Ba, La, Nd and Sm 
are present (see Fig. \ref{late_s}). For early-R stars we also used the Ba I-II lines at $\lambda~6498.9$ and $\lambda~6496.9$ \AA{}, 
respectively. We attempted to detect the radioactive element technetium from the resonant lines at $\lambda~4260$ \AA{}
and the recombination Tc I line at $\lambda~5924.5$ \AA{}. We fail to detect the bluer line in all the
stars mainly because of the low S/N ratio achieved in the spectra in that region. Upper limits on the Tc abundance were set
in the stars HIP 39\,118 and HIP 109\,158 by using the recombination line. This result is compatible with the presence of some
$s$-element enhancement (see Table \ref{sabundances}) in these two stars. However, because we can barely reproduce the blend
at $\lambda$ 5924 \AA{}, these Tc upper limits have to be taken with caution. Finally, Rb abundances were determined using the Rb I line at $\lambda~7800.2$ \AA.  In this region there are two Ni I lines at $\lambda$ 7788.9, 7797.6 \AA{} 
and three Fe I lines at $\lambda\lambda 7780.6, 7802.5$, and 7807.9 \AA{} not very much blended by
molecular absorptions that can be used to derive the average metallicity.
\setlength\tabcolsep{0.25cm}
\begin{table*}
\begin{center}
\caption{Dependence of the derived abundances on the atmospheric parameters}
\label{errors}
\normalsize{
\begin{tabular}{c cc c c cc cc}
\hline\hline  
 Species &$T\rm{_{eff}}^{~a}$&$\rm \log$ g&$\rm [M/H]$&$\rm \xi$&$\rm macro.$
& $\rm C/O$ & $\rm ^{12}C/^{13}C$&Total\\
X&$\pm$150/250K&$\pm$0.5&$\pm$0.2&$\pm$0.5 km s$^{-1}$&$\pm$1.5 km s$^{-1}$
&$\pm$0.1 & $\pm$10\\
\hline
\multicolumn{9}{c}{Late-R stars}\\
\hline
$\rm Li$ &	$\pm$0.40	&	$\pm$0.05	&	$\pm$0.20	&	$\mp$0.00	&	$\pm$0.00	&	$\mp$0.20	&	$\pm$0.00&$\pm$0.50			\\
$\rm Fe $ &	$\pm$0.30	&	$\pm$0.00	&	-	&	$\mp$0.05	&	$\pm$0.10	       &	$\mp$0.10	&	$\pm$0.00&$\pm$0.35			\\
$\rm Rb $ &	$\pm$0.20	&	$\pm$0.10	&	$\pm$0.20	&	$\mp$0.00	&	$\pm$0.00	&	$\mp$0.15	&	$\mp$0.10&$\pm$0.35			\\
$\rm Sr $ &	$\pm$0.15	&	$\pm$0.00	&	$\pm$0.10	&	$\mp$0.05	&	$\pm$0.20	&	$\mp$0.10	&	$\mp$0.05&$\pm$0.30		\\
$\rm Y $ &	$\pm$0.15	&	$\pm$0.00	&	$\pm$0.10	&	$\mp$0.20	&	$\pm$0.25	&	$\mp$0.05	&	$\pm$0.00&$\pm$0.35			\\
$\rm Zr $ &	$\pm$0.05	&	$\pm$0.00	&	$\pm$0.05	&	$\mp$0.20	&	$\pm$0.25	&	$\mp$0.00	&	$\pm$0.00&$\pm$0.35			\\
$\rm Ba $ &	$\pm$0.15	&	$\pm$0.10	&	$\pm$0.10	&	$\mp$0.05	&	$\pm$0.00	&	$\pm$0.00	&	$\pm$0.00&$\pm$0.25			\\
$\rm La $ &	$\pm$0.05	&	$\pm$0.20	&	$\pm$0.10	&	$\mp$0.10	&	$\pm$0.20	&	$\mp$0.05	&	$\pm$0.00&$\pm$0.35			\\
$\rm Nd $ &	$\pm$0.05	&	$\pm$0.20	&	$\pm$0.05	&	$\mp$0.20	&	$\pm$0.20	&	$\mp$0.05	&	$\pm$0.00&$\pm$0.35			\\
$\rm Sm $&	$\pm$0.05	&	$\pm$0.10	&	$\pm$0.10	&	$\mp$0.25	&	$\pm$0.20	&	$\mp$0.00	&	$\pm$0.00&$\pm$0.35			\\
\hline
\multicolumn{9}{c}{Early-R stars}\\
\hline
$\rm Li $     &	$\pm$0.10	&	$\pm$0.00	&	$\pm$0.10	&	$\mp$0.00	&	$\pm$0.00	&	$\mp$0.10       &	$\pm$0.05&$\pm$0.30	\\
$\rm Fe $     &	$\pm$0.05	&	$\pm$0.00	&	-	&	$\mp$0.05	&	$\pm$0.15	&	$\mp$0.10	&	$\pm$0.00&$\pm$0.20			\\
$\rm Rb $	&	$\pm$0.05	&	$\pm$0.00	&	$\pm$0.00	&	$\mp$0.00	&	$\pm$0.05	&	$\mp$0.05	&	$\pm$0.00&$\pm$0.20	\\
$\rm Sr $	&	$\pm$0.10	&	$\pm$0.00	&	$\mp$0.05	&	$\mp$0.10	&	$\pm$0.15	&	$\mp$0.05	&	$\mp$0.00&$\pm$0.20		\\
$\rm Y $	&	$\pm$0.10	&	$\pm$0.00	&	$\pm$0.25	&	$\mp$0.00	&	$\pm$0.10      &	$\mp$0.15	&	$\pm$0.00&$\pm$0.40	\\
$\rm Zr$	&	$\pm$0.10	&	$\pm$0.00	&	$\pm$0.00	&	$\mp$0.00	&	$\pm$0.10	&	$\mp$0.05	&	$\pm$0.00&$\pm$0.30		\\
$\rm Ba$	&	$\pm$0.10	&	$\pm$0.10	&	$\pm$0.20	&	$\mp$0.00	&	$\pm$0.10	&	$\mp$0.05	&	$\pm$0.00&$\pm$0.30		\\
$\rm La $	&	$\pm$0.05	&	$\pm$0.20	&	$\pm$0.10	&	$\mp$0.00	&	$\pm$0.15	&	$\mp$0.05	&	$\mp$0.00&$\pm$0.30	\\
$\rm Nd $	&	$\pm$0.15	&	$\pm$0.20	&	$\pm$0.05	&	$\mp$0.05	&	$\pm$0.20	&	$\mp$0.00	&	$\mp$0.00&$\pm$0.40	\\
$\rm Sm $     &	$\pm$0.15	&	$\pm$0.20	&	$\mp$0.10	&	$\mp$0.00	&	$\pm$0.15	&	$\mp$0.00	&	$\mp$0.00&$\pm$0.35	\\
\noalign{\smallskip}
\hline
\end{tabular}
}
\end{center}
\scriptsize{$^a$ See text for details.}
\end{table*}

Table \ref{errors} shows the uncertainty in the abundances derived for the different chemical species due
to the uncertainty in the model atmosphere parameters. To do this a canonical model
atmosphere with parameters $T\rm{_{eff}}$/log g/[M/H]=  4750/2.0/0.0  and 3300/0.0/0.0 was used for early- and late-R
stars, respectively. Due to the very different sensitivity of the theoretical spectra to changes in the effective
temperature, changes in the abundances due to this parameter were calculated assuming variations 
by $\pm 150$ and $\pm 250$ K for early- and late-R stars, respectively.

\setlength\tabcolsep{0.25cm}
\begin{table*}
\begin{center}
\caption{Abundances derived and relevant abundances ratios.}
\label{labun}
\normalsize{
\begin{tabular}{c c c c c c c c c c c}
\noalign{\smallskip}                                                                                 
\hline\hline  
Name&C/O &$^{12}$C/$^{13}$C&Li&Mn&Fe&Ni&Zn&$\rm{[M/H]}$&[C/M]&[N/M]\\
\hline\noalign{\smallskip}
\multicolumn{11}{c}{Late-R stars}\\
 \hline\noalign{\smallskip}
HIP 35\,810 &1.12  &  65 & 0.20    &-&7.07  &5.93 &4.20& $-$0.38   &0.39&-   \\
HIP 36\,623 &1.10  &  23 & $-$0.20&-&7.18  &5.92 &-& $-$0.27      &0.23&0.19\\
HIP 62\,401 & -    & -             &-&-     &-&-&-&-       &  -    & -     \\
HIP 91\,929 &1.05  &  58 & $-$0.50 &-&7.45  &6.23 &-& 0.00         &0.29&-       \\
HIP 108\,205&1.02  &  90 &$-$1.50  &-&7.45  &- &-& 0.00            &0.28&-      \\
HIP 109\,158&1.05  &  85 & $-$1.00&-&7.43  &5.93 &-& $-$0.02      &0.31&-    \\
\noalign{\smallskip}\hline
\multicolumn{11}{c}{Early-R stars}\\
\hline\noalign{\smallskip}
HIP 39\,118&$>$0.50 &13:& 0.85   &5.10 &7.16&5.93&-&$-$0.29    &0.67&0.51       \\
HIP 44\,812&1.23&5  & $<$1.00&-    &7.42&6.23&4.60&$-$0.03 &0.39&0.95           \\
HIP 53\,832&1.05&24 & $<$0.60&4.69 &6.68&5.43&3.80&$-$0.77 &0.46&1.09         \\
HIP 58\,786&1.45&70 & $<$1.00&-    &7.16&5.90&4.15&$-$0.29 &0.53&0.11          \\
HIP 62\,944&0.56&22 & 2.60   &5.49 &7.57&6.28&4.70& 0.12  &0.29&0.40            \\
HIP 69\,089&0.87&19 & 1.80   &5.25 &7.28&6.04&4.50&$-$0.17 &0.18&0.79         \\       
HIP 74\,826&1.17&20 &$<$0.50 &5.10 &7.15&5.95&4.30&$-$0.30 &0.34&0.70        \\
HIP 82\,184&0.81&10 &$<$0.40 &5.24 &7.30&6.08&4.45&$-$0.15 &0.18&0.62        \\
HIP 84\,266&1.65&7  &$<$1.05 &5.29 &7.35&6.13&4.50&$-$0.10 &0.49&0.72       \\
HIP 85\,750&1.05&22 &$<$0.30 &4.89 &6.97&6.97&$<$3.85&$-$0.48 &0.47&0.68        \\
HIP 86\,927&0.65&7  &$<$0.48 &5.35 &7.40&7.40&$<$4.55&$-$0.05 &0.13&0.25       \\
HIP 87\,603&1.74&9  &$<$1.10 &-    &6.93&6.93&4.20&$-$0.52 &0.73&0.42       \\
HIP 88\,887&1.38&5  &$<$0.50 &-    &7.36&7.36&4.30&$-$0.09 &0.50&1.01       \\
HIP 94\,049&1.32&9  &$<$0.40 &-    &6.83&6.83&-&$-$0.62 &0.71&0.82       \\
HIP 95\,422&2.34&6  &$<$0.50 &-    &7.19&7.19&-&$-$0.26&0.60&-           \\
HIP 98\,223&1.20&16 &$<$0.00 &-    &6.66&6.66&-&$-$0.79&0.45&0.11         \\
HIP 113\,150&3.09&9 &$<$0.55 &-    &6.98&6.98&-&$-$0.47&0.93&-            \\
\hline     
\end{tabular}
}
\end{center}
\scriptsize{: indicates uncertain value.}
\end{table*}

%\begin{landscape}
%\addtolength{\evensidemargin}{-1.0cm}{
\begin{table*}
\setlength\tabcolsep{0.13cm}
\begin{center}
%\centering
\caption{Heavy element abundance ratios.}
\label{sabundances}
\normalsize{
\begin{tabular}{crrrrrrrrrrrr}
\noalign{\smallskip}
\hline\hline\noalign{\smallskip}  
Name &[Rb/M] &[Sr/M]&[Y/M]&[Zr/M]&[Ba/M]&[La/M]&[Nd/M]&[Sm/M]&[ls/M]&[hs/M]&[hs/ls]&[s/M] \\
\hline\noalign{\smallskip}
\multicolumn{13}{c}{Late-R stars}\\
\hline\noalign{\smallskip}
HIP 35\,810 &  -     & 0.26&0.27&0.29  &  0.78   & 0.55 & 0.83 &0.17 & 0.27    & 0.58     & 0.31 &0.43                        \\
HIP 36\,623 & 0.10   & 0.95&1.31   &1.13  &  1.72   & 1.64 & 1.72 &0.56 & 1.13    & 1.41     & 0.28 &1.27                        \\
HIP 91\,929 &  -     & -    &$-$0.01&0.01 &  $-$0.02&  -   &   -  &  -  & 0.00    & $-$0.02    & $-$0.02 &$-$0.01    \\
HIP 108\,205&  -     & -    &0.59   & -   &  0.73   &  -   &   -  &  -  & 0.59    & 0.73     & 0.14 &0.66            \\ 
HIP 109\,158&  -     & -    &0.81   & -   &  0.85   &  -   &   -  &  -  & 0.81    & 0.85     & 0.04 &0.83            \\
\noalign{\smallskip}
\hline
\multicolumn{13}{c}{Early-R stars}\\
\hline\noalign{\smallskip}
HIP 39\,118 & 0.30    & 0.22  & 0.68   &  0.70  & 0.49  &  0.46  & 0.64  &  0.58 & 0.53	&0.54	&0.01   &0.54   \\
HIP 44\,812 &  -      & $-$0.07 & 0.03   &  $-$0.26 & $-$0.04 &  0.13  &   -   &  0.12 & $-$0.10 &0.07   &0.17 &$-$0.01   \\
HIP 53\,832 & $<$0.70    & 0.85  & 1.06   &  0.78  & 1.65  &  $<$2.04  & 1.42  &  1.36 & 0.90	&1.62	&0.72 &1.26   \\
HIP 58\,786 &  -      & 0.09  & $-$0.06  &  $-$0.30 & 0.02  &  $-$0.01 &   -   &  $-$0.01& $-$0.09 &0.00	&0.09 &$-$0.04     \\ 
HIP 62\,944 & $-$0.09   & $-$0.02 & $-$0.17  &  $-$0.41 & $-$0.29 &  $-$0.02 & $-$0.12 &  0.07 & $-$0.20 &0.09   &0.11  &$-$0.15  \\
HIP 69\,089 & $-$0.13   & 0.05  & 0.16   &  $-$0.12 & 0.00  &  0.07  & 0.07  &  0.06 & 0.03	&0.05	&0.02 &0.04  \\
HIP 74\,826 & 0.01    & $-$0.02 & 0.02   &  0.01  & 0.08  &  $-$0.04 & 0.00 &  $-$0.01& 0.00	&0.01	&0.00 &0.01                    \\
HIP 82\,184 & $-$0.01   & 0.00  & $-$0.07  &  $-$0.14 & 0.08  &  0.15  & $-$0.10 &  0.00& $-$0.07 &0.03   &0.10 &$-$0.02             \\
HIP 84\,266 & $-$0.04   & 0.00 & $-$0.21  &  $-$0.11 & $-$0.07 &  0.22  & 0.05  &  0.14 & $-$0.11 &0.09   &0.19 &$-$0.01            \\
HIP 85\,750 & $-$0.05   & 0.76  & 0.97   &  0.59  & 1.56  &  1.65:  & 1.53  &  1.27 & 0.77	&1.45	&0.68 &1.11  \\
HIP 86\,927 & 0.18    & 0.00  & $-$0.01  &  0.31  & 0.20  &  0.32  & 0.40  &  0.19 & 0.10	&0.28	&0.18 &0.19  \\
HIP 87\,603 & 0.05    & 0.12  & 0.11   &  $-$0.07 & 0.45  &  0.52  & 0.12  &  0.11 & 0.05	&0.30	&0.25  &0.18      \\
HIP 88\,887 & $-$0.07   & $-$0.13 & $-$0.02  &  $-$0.20 & $-$0.18 &  $-$0.11 & $-$0.11 &  $-$0.12& $-$0.12 &$-$0.13	&$-$0.01&$-$0.12 \\
HIP 94\,049 &  -      & -     & 0.02   &   -    & $-$0.05 &   -    &   -   &   -   & 0.02	&$-$0.05 &$-$0.07 &$-$0.01 \\
HIP 95\,422 &  -      & -     & $-$0.04  &   -    & 0.11  &   -    &   -   &    -  & $-$0.04 &0.11   &0.15 &0.04    \\
HIP 98\,223 &  -      & 0.67  & 1.13   &  1.00  & 1.25  &   -    & 1.44  &  1.18 & 0.93	&1.29	&0.36 &1.11  \\
HIP 113\,150&  -      & -     & $-$0.03  &  $-$0.02 & 0.00  &   -    &   -   &  $-$0.04& $-$0.02&$-$0.02	&0.00 &$-$0.02   \\
\noalign{\smallskip}
\hline
\noalign{\smallskip}
\end{tabular}
}
\end{center}
\scriptsize{ls $= \langle$Sr, Y, Zr$\rangle$, hs $= \langle$Ba, La, Nd, Sm$\rangle$ and s$ = \langle$ls $+$ hs $\rangle$.\\
 }
\end{table*}
%}

The final uncertainty in the abundance is found by considering quadratically all the uncertainties including the
uncertainty in the continuum position ($< 5\%$) and, when more than 3 lines of a given element were used, the abundance
dispersion among the lines of this element as an additional source of uncertainty. This is indicated in the last
column of Table \ref{errors}. From this table it can be appreciated that the error in the absolute abundances derived are in the range 
$\rm\Delta [X/H]=\pm (0.2-0.5)$ dex, being larger for late-R stars. When considering the abundances ratios ([X/Fe]) relative
to iron, the error is reduced if the variation on a given atmospheric parameter affects in the same way 
the element of interest. The results of the chemical analysis are summarised in Tables \ref{labun} and \ref{sabundances}.
The last four columns in Table \ref{sabundances} show the average enhancement of light $s$-elements, ls$ = \langle$Sr, Y, Zr$\rangle$, 
heavy $s$-elements, hs$ = \langle$Ba, La, Nd, Sm$\rangle$, the $s$-element index [hs/ls], and the total average 
$s$-element enhancement.

\section{Results and discussion} \label{results}

\subsection{Comments on particular stars} \label{comments}

The detailed chemical composition reported in Tables \ref{labun} and \ref{sabundances} shows
that there are several stars that do not fulfil one of the main characteristic that defines
a R star, i.e., a carbon star without $s$-element enhancements. HIP 39\,118 is one of these stars, showing a
clear $s$-element enhancement, [s/M] $\sim +0.54$.  In fact, our best fit to the Tc I line 
at $\lambda$ 5924 \AA{} in this star gives log $\epsilon$(Tc)$ < 1.20$. This, together with the
$s$-element enhancement found would be an indication of this star being on the AGB phase. However considering that the luminosity 
of this star is too low to be on the AGB phase (see Table \ref{magnitudes}), and the difficulty in reproducing the $\lambda$ 5924 \AA{} Tc
blend, this Tc upper limit has to be considered with extreme caution and thus, it would be risky to conclude
that HIP 39\,118 is an AGB star. HIP 39\,118 has other chemical peculiarities. It is not a carbon star since its C/O ratio ($\sim 0.50$) is considerably
lower than unity even considering the uncertainty in the derivation of this ratio. Indeed, the $\lambda~6300$ {\AA}
[OI] line is very strong in this star which indicates some O enhancement, [O/Fe]$<0.68$. Titanium, another alpha
element, is also enhanced in this star [Ti/Fe]$=+0.3$. Due to its low luminosity (M$\rm_{bol}= 0.47$)
and $s$-element enhancement one is tempted to ascribe this star as a classical barium star
that are known to be all binaries. In the case of HIP 39\,118, there are only two radial velocity measurements within a 5 days interval 
separation \citep{platais2003}. With such short time span compared with the long orbital periods of the classical barium stars, no conclusion
may be drawn regarding binarity. Moreover, the moderate 
Li abundance in this star, log $\epsilon$ (Li)$= 0.85$, is difficult to explain in a mass transfer scenario.
Thus, the spectral classification of this star is uncertain and we just put it in the group of chemically anomalous
red giant stars, but certainly not an early-R star.

HIP 53\,832 has an abundance pattern typical of a classic CH-type star:
a moderate metal-poor carbon-rich ([M/H] $= -0.77$, C/O $= 1.05$) star with a relatively low carbon isotopic ratio,
($^{12}$C/$^{13}$C$ = 24$) and an important $s$-element enhancement ([s/M]$ = +1.26$). In fact, the binary nature of this star 
has been confirmed by radial velocity measurements by \citet{platais2003} and thus, there is no doubt about its
classification as a CH star. 

HIP 62\,944 is one of the first Li-rich giants discovered \citep{wallerstein1982}. Our Li abundance in this
star, log $\epsilon$(Li)$ = 2.60$, agrees with that deduced by these authors. The CNO content and the metallicity 
derived here are also in agreement. Again this star is not a carbon star as the C/O ratio is near the solar value
(see Table \ref{labun}). The low isotopic ratio ($^{12}$C/$^{13}$C$ = 22$) derived here is similar to those found in
others Li-rich giants \citep{delareza2006}. The origin of Li in RGB stars is not well understood. Some kind of non-standard mixing mecha\-nism
du\-ring the RGB phase \citep[see e.g.][]{palacios2001,palacios2006,guandalini2009} seems to be needed. 
In any case, HIP 62\,944 is a K giant, not a R star.

HIP 69\,089 presents some C enrichment although its C/O ratio is below unity (0.87, see Table \ref{labun}). 
As in the case of HIP 62\,944, this star is also Li-rich (log $\epsilon$(Li)$=1.80$) with a low 
carbon isotopic ratio (19) typical in these stars. The Li abundance previously derived by \citet{luck1995} 
in this star, log $\epsilon$(Li)$ = 2.04$, is compatible with the value
derived in the present work. The $s$-element abundances are solar scaled. On the basis of these values
we classify it as another Li-rich K giant.

We derive significant $s$-element overabundances in the carbon star HIP 85\,750 (see Table \ref{sabundances}). Radial velocity variations have been detected 
in this star (Table \ref{obs}). Since it is a moderately metal-poor star, 
we classify it as a CH-type star.

According to our analysis HIP 86\,927 is not a carbon star (C/O$ = 0.65$), and shows no $s$-element enhancement. 
Its carbon isotopic ratio is low (7) and agrees with the previous determination by \citet{dominy1984}. Unfortunately, this
author did not derive any other abundance signature in this star. Further, the luminosity of this star
(M$\rm_{bol} = 0.22$) is too low for a typical early-R star (see Sect. \ref{luminosity}), thus we believe that its characteristics 
are closer to a normal K giant which has undergone some extra-mixing revealed by its low $^{12}$C/$^{13}$C ratio.

Finally, HIP 98\,223 is a metal-poor ([M/H] = $-0.79$) carbon star showing 
a large enrichment in $s$-elements ([s/M]$ = +1.11$). These are the typical abundance signatures of the CH stars.

Among the late-R stars we have two {\it peculiar} stars. 
HIP 36\,623 is a symbiotic star \citep{johnson1988,belczynski2000,munari2002} whose luminosity (see Table \ref{magnitudes}) 
and chemical composition (see Tables \ref{labun} and \ref{sabundances}) are compatible with those expected at the AGB phase. 
Perhaps the envelope of this star has undergone two periods of mixing: the first one by the material accreted from the
primary star, and the second one, a self-contamination triggered by third dredge-up
episodes on the AGB phase. Thus, it is proba\-bly a normal N-type star. 
The spectrum of this star looks similar to the other late-R in the sample without any evidence of contamination by
emission lines or emission continua coming from the hot UV radiation of the companion (probably a white dwarf). The broad
emission features around 6825 and 7082 {\AA} (due to O VI photons Raman scattered off H I) are not present either. These features
are found in some symbiotic stars (e.g. AG Dra, \citealt{smith1996}). Thus, we believe that the abundances of the $s$-elements
in this star (mostly derived from the blue part of the spectrum) are not underestimated because of a contribution to the
spectral continuum by the companion star. On the other hand, we do not find
$s$-element overabundances in the late-R HIP 91\,929. However its spectrum is identical to those of N-type stars.
In Paper II we found a few N-type stars where the $s$-element content was compatible with no enhancements
within the error bars. Perhaps HIP 91\,929 is another example of this. Maybe these stars are at the beginning of the
thermally pulsing AGB phase (as their luminosity might indicate) and did not have time yet to pollute the envelope
with $s$-elements.

In summary, out of seventeen stars classified as early-R stars we have found seven stars ($\sim$ 40 $\%$)
that are not R stars. Since the stars in our sample were randomly chosen from the Hipparcos catalogue (the only 
condition was to be observable from the north hemisphere) and although the sample is limited, we confirm the
previous claims that among the R stars there is a significant number of stars which are wrongly classified. This is
an important result since it might discard the previous belief (see Sect. \ref{intro}) that the R stars (namely early-R) 
represent a frequent stage in the evolution of low-mass stars.

\subsection{C/O and $^{12}$C/~$^{13}$C ratios}
\begin{figure}[ht!]
\centering
\includegraphics[width=9cm,angle=0]{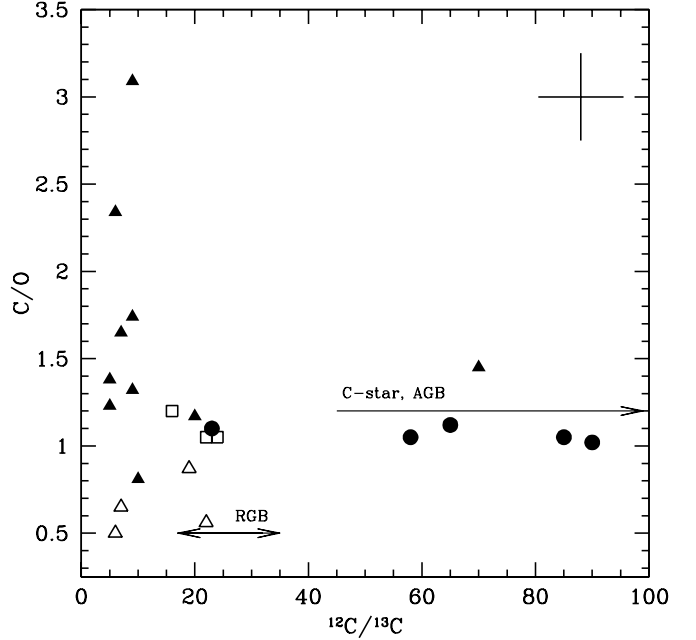}
\caption{C/O vs. $^{12}$C/$^{13}$C ratios. Symbos as Fig.2. 
The arrows indicate the expected range in the $^{12}$C/$^{13}$C ratio
according to the standard evolutionary models of low-mass stars in the RGB and AGB phases (see text for
details).}
\label{ciso_co}
\end{figure}

Fig. \ref{ciso_co} shows the C/O and $^{12}$C/$^{13}$C ratios found in our stars (C$=^{12}$C$+^{13}$C).
The stars reclassified here (see previous section) are indicated by open symbols (squares, CH stars and triangles, K giants)
In the figure are indicated the expected values of the $^{12}$C/$^{13}$C ratio according to 
standard low-mass stellar models \citep[e.g.][]{cristallo2009} after the first dredge-up, and the AGB phase, 
respectively. The expected range in the AGB phase has been computed considering 
the effect of an extra-mixing mecha\-nism after the first dredge-up \citep{booth1999}, which would reduce further
the $^{12}$C/$^{13}$C ratio to a typical value of $\sim$ 12. Such $^{12}$C/$^{13}$C ratios are frequently found in 
Population I giant stars \citep{gilroy1989,gilroy1991} and can be only explained assuming a
non standard mixing process \citep{charbo1998}. It is evident from Fig. \ref{ciso_co} that most of the {\it true} 
early-R stars have carbon isotopic ratios far below the minimum values expected after the first dredge-up (RGB phase)  
and the thermally pulsing AGB phase. Further, since early-R stars show also N enhancements (see Table \ref{labun}, and
\citealt{dominy1984}), the low $^{12}$C/$^{13}$C ratios together with the N overabundances ($<$[N/M]$>=+0.6$) would indicate that the material 
in the envelope of these stars has been processed through the CNO bi-cycle (almost at the equilibrium).
 If that were the case, material exposed to the CNO bi-cycle is expected to be depleted in $^{16}$O by a 
factor of at least $\sim 40$, the exact value depending on the burning temperature.  
Even considering the large uncertainty in the O abundance in our stars (actually we scale the O abundance with the average metallicity,  see Sect. 3), we can safely discard a large O depletion from the analysis: the C/O ratios would be much higher than the values found here and because few C would be locked into CO, the C-bearing molecules (CN, CH and C$_2$) would appear very strong in the spectrum which is not observed in any of the early-R stars. 
Furthermore, for the few early-R stars where O abundance has been determined \citep{dominy1984} no evidence of oxygen 
depletion is found. This fact casts some doubt on the idea that the mixing of CNO-equilibrium material is the only responsible for the early-R stars. 
The C and N enhancements found are rather evidence that the material that we currently observe in the envelopes of early-R stars has been processed by both 
the CN cycle and He-burning. Probably, C-rich material was mixed with protons on a time scale short enough and at a temperature 
low enough (T$\leq 7\times 10^7$ K) to allow only the reactions $^{12}$C$(p,\gamma)^{13}$C and $^{13}$C$(p,\gamma)^{14}$N, without any significant operation 
of the ON-cycle. In any case, the determination of the oxygen isotopic ratios in early-R stars would be a valuable tool to evaluate the role played by 
the CNO bi-cycle in the surface composition of these stars.

On the other hand, the $^{12}$C/$^{13}$C ratios derived in late-R stars are significantly higher
than those for the early ones. In these stars observations in general agree with AGB model predictions
(see Fig. \ref{ciso_co}). An exception is the star HIP 36\,623 with a $^{12}$C/$^{13}$C ratio below the predicted values 
for the AGB carbon stars. However, these low  $^{12}$C/$^{13}$C values have been found also in many 
N-type stars \citep{ohnaka1996,abia2002}  and are currently interpreted as an evidence of the operation 
of a non standard mixing process also in low-mass stars during the AGB phase. Differences between the early- and late- 
R stars are also evident in the C/O ratio. Early-R stars show a wider range in the C/O ratio while late-R stars are concentrated 
very close to 1, as in the galactic N-type stars
of similar metallicity \citep{lambert1986,abia2002}. This is consistent with late-R stars having
higher mass than early-R stars because it is more difficult to increase the C/O ratio of a more massive star.

\begin{figure}
\centering
\includegraphics[width=9cm,angle=0]{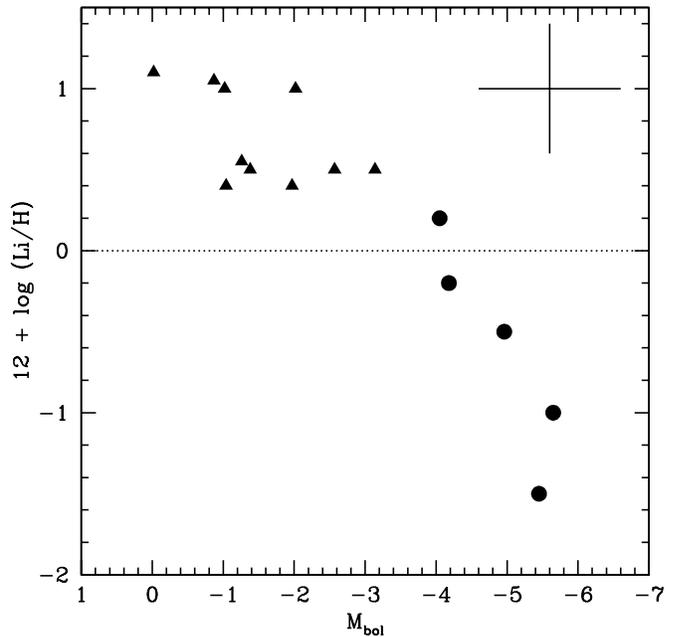}
\caption{Derived Li abundances and the absolute bolometric magnitude according to Bergeat et al. (2002a).
Symbols as in Fig. 2.}
\label{li}
\end{figure}

\subsection{Lithium}

As noted in Sect. \ref{estimation}, most of the Li abundances derived in early-R stars should be considered upper limits. In cooler
late-R stars, synthetic spectra are much more sensitive to changes in the Li abundance (but note however the large formal 
uncertainty in the Li abundance, see Table 4). Fig. \ref{li} presents the Li abundances derived in late- and early-R stars (excluding 
the stars reclassified in the
previous section) and the luminosity. We remind that the typical stellar mass of early- and late-R stars is probably very different 
(see Sect. \ref{distri}) thus, one should not establish a sequence of Li depletion in R stars from this figure.
Instead, the apparent Li abundance vs. luminosity relationship have to be interpreted separately. By
comparing stars of the same spectral type we found that early-R stars present very similar Li abundances 
(in the range log $\epsilon$(Li)$\sim 0.5-1.0$) without any evident correlation with the luminosity. These upper limits in the 
Li abundances are higher than the expected va\-lues in post-tip RGB stars (marked with the dotted line at 
log $\epsilon$(Li)$=0.0$ in Fig. 7, e.g. \citealt{castilho2000}). Although, it would be necessary to confirm the possible Li detection with higher 
resolution spectra, this is of some relevance because it may constrain the scenarios proposed to form a R star. 

In late-R stars, at contrary, Li abundances decrease with the increasing luminosity as one would expect to occur as the stellar envelope goes deeply 
toward the interior of the star during the ascend along the AGB phase. Indeed, the Li abundances derived in our late-R stars are similar 
to those derived in normal AGB carbon stars \citep{denn1991,abia1993}, which supports again the idea that they are identical to the N-type stars.

\subsection{Abundances of heavy elements}

\begin{figure}
\centering
\includegraphics[width=9.5cm,angle=0]{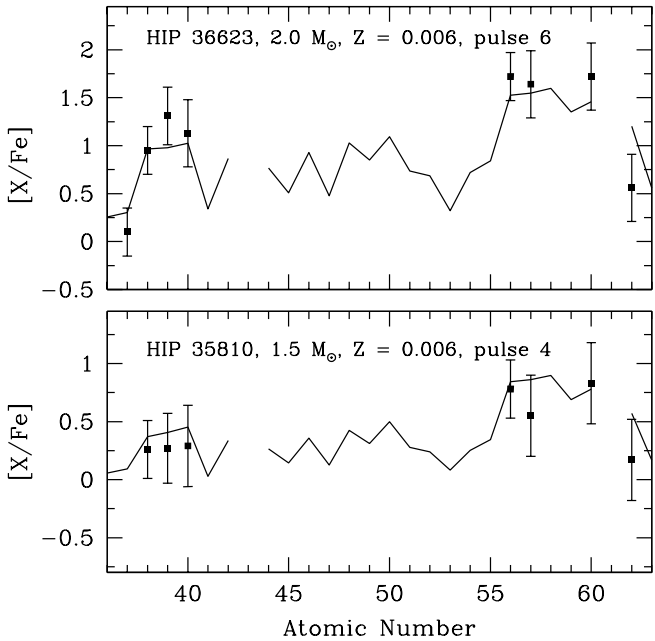}
\caption{Comparisons of the derived $s$-element ratios ([X/Fe]) in the late-R stars HIP 36\,623 and
HIP 35\,810 with the theoretical predictions by \citet{cristallo2009} for two different low-mass
AGB models. In each panel is indicated the stellar mass, metallicity and specific thermal pulse
that provide the best fit to the observed abundances.}
\label{spattern_late}
\end{figure}

%\begin{figure}
%\centering
%\includegraphics[width=9.5cm,angle=0]{selem_ch.eps}
%\caption{Reproduction of the observed [X/Fe] ratios in the CH-type stars and in HIP 39\,118 (extrinsic cases).
%It is indicated the mass, metallicity and dilution factor
%that provide the best fit to the abundances derived in the chemical analysis. This dilution factor is
%computed considering the abundances in the last thermal pulse, using \citet{cristallo2009} low-mass AGB models.}
%\label{spattern_ch}
%\end{figure}  

Table \ref{sabundances} summarises the heavy element abundances derived in our stars. In the sample of early-R stars, and
excluding those stars that we have reclassified as belonging to other spectral types (see above), it is evident that
these stars do not have $s$-element overabundances [X/M]$\sim 0.0$ within the
error bars. This is the case of HIP 44\,812, HIP 58\,786, HIP 74\,826, HIP 82\,184, HIP 84\,266, HIP 87\,603, HIP 88\,887, HIP 94\,049, 
HIP 95\,422 and HIP 113\,150. Thus, we confirm the analysis by \citet{dominy1984} in this kind of stars.
Late-R stars do have $s$-element enhancements\footnote{The late-R star HIP 62\,401 is excluded from the chemical
analysis because its spectrum shows very broad lines that we cannot reproduce. This is typically found in Mira
variables, see Table \ref{obs}.} at the same level than those found in galactic N-type stars of similar metallicity, $<$[s/M]$>= 0.64$ \citep{abia2002}. 
In fact, we detect Tc (using the weak recombination line at $\lambda$ 5924.47 \AA)
in HIP 109\,158 and it may be also present in HIP 108\,205. This is usually interpreted as evidence of the {\it in situ}
production of the $s$-elements in the star, the envelope being polluted by the operation of the
third dredge-up during the AGB phase. These stars are named {\it intrinsic}. Fig. \ref{spattern_late} shows an example
of a detailed reproduction of the observed heavy element pattern in two late-R stars by theoretical $s$-process nucleosynthesis  calculations in low-mass 
AGB stars. (see details in \citealt{cristallo2009}). Each panel
indicates the stellar mass model and thermal pulse giving the best fit to the observed pattern. The metallicity
in the model ($Z$) was chosen as close as possible to the observed metallicity in the stars. By selecting a stellar mass
model, metallicity and specific thermal pulse, fits of similar quality can be found for the other late-R stars
in our sample, as well as for the early-R stars reclassified as CH-type (HIP 98\,223, HIP 85\,750 and HIP 53\,832) which also
show heavy element enhancements. In this case, we can mimic their extrinsic nature (i.e. the chemical composition
of the envelope is due to the pollution by the matter accreted from a companion) including an extra-parameter in the
model: the dilution factor, defined as the ratio between the mass of the envelope and the mass transferred from the
primary star (e.g. \citealt{bisterzo2006}). The dilution factor needed are in the range  $2-3$. These values are consistent with a secondary star having 
a large envelope, most of the CH-type star are indeed giants.

We  close this section realising once more that the most pro\-bable neutron source producing the observed $s$-element
pattern in these stars is the  $^{13}$C$(\alpha,n)^{16}$O reaction which is found the main neutron donor in low-mass
AGB stars \citep{lambert1995,abia2001}. This is concluded from the observed 
[Rb/Sr,Y,Zr]$\leq 0$ (see Table \ref{sabundances}). In AGB stars of intermediate mass (M $\geq 3$ M$_\odot$) probably 
dominates the $^{22}$Ne$(\alpha,n)^{25}$Mg neutron source resulting in higher neutron densities  (N$_n\geq 10^{10}$ cm$^{-3}$) and
so [Rb/Sr,Y,Zr]$>0$ (e.g. \citealt{garcia2006}).

\subsection{Evolutionary status}

 We have shown in the previous analysis that the true early- and late-R stars have
different properties. Most of the late-R stars are long-period variables of the SR or
Mira types (see Table \ref{obs}), and they often exhibit excess emission at 12 $\mu$m due to dust,
indicative of mass loss. In the colour-colour diagram they occupy the same location as N-type
stars and have similar luminosities (or slightly lower). We have also shown that
there is no significant difference in the chemical composition of late-R stars and N-type
stars, both types showing $s$-element enhancements at the same level. They are thus closely related, 
or are identical. Definitely late-R stars are AGB stars and, most probably, mark the bottom of 
the AGB phase. 

Concerning early-R stars, our study (although limited in the number of stars) confirms previous claims
that most carbon stars catalogues contain a significant number of stars wrongly classified as 
early-R stars. In a sample of seventeen early-R stars we have found seven ($\sim 40\%$) objects that
are K giants or CH-type stars. This finding, together with the definite ascription of late-R stars 
as N stars certainly reduce the number of real (early) R type stars among all the giant carbon stars. 
Therefore, R stars are indeed scarce objects, but we still lack an evolutionary scenario to explain
their observational properties. We have to consider at least three possibilities: i) mass transfer in a binary system, 
ii) original pollution, and iii) a non-standard mixing mechanism able to mix carbon
to the surface. Due to their position in the H-R diagram, and their peculiar chemical composition, the favoured 
scenario is the later one \citep{dominy1984,knapp2001} triggered by an anomalous He-flash (e.g. \citealt{mengel1976,pacinski1977}).

As noted in Sect. 1, all early-R stars seem to be single stars: so far, attemps to detect radial velocity variations due
to binarity have failed  \citep{mcclure1997}. This is indeed a very improbable figure since in any
stellar population one would expect a minimum of $\sim 30\%$ binary systems. McClure (1997) used this argument
to suggest that R stars were initially all binaries and that they coalesced into a single object during their evolution. In this
hypothesis one might argue that the carbon enrichment that we currently observe was a
consequence of mass transfer prior to the coalescence. However, it is extremely difficult to form carbon stars from mass transfer at 
near solar metallicity (\citealt{abia2003,masseron2009}, although a few exceptions seem to exist, e.g., BD +57$\degr$ 2161, 
see \citealt{zacs2005}), thus there is very little or not space for the mass transfer scenario.
The se\-cond scenario (carbon excess in the original gas cloud) it also ruled out because one should expect to find carbon stars of near 
solar metallicity at earlier evolutionary stages (main sequence, turn-off, sub-giants stars etc.). This is 
the case of the carbon rich metal-poor stars discovered in the galactic halo (e.g. \citealt{aoki2007,masseron2009}): they are found along all the H-R diagram from the main sequence to the RGB. However, no a sole star with similar charac\-teristics to those of early-R has been found in earlier evolutionary stages. In the same line, we might ask where are 
the descendants or early-R stars? Once leaving the He-core burning phase, they will evolve and ascend 
the giant branch for a second time, i.e. they should become AGB stars. The carbon stars of J-type have been traditionally 
proposed  as the {\it daughters} of early-R stars \citep[e.g.][]{lloyd1986} because of their 
chemical similarities (both are carbon stars without $s$-element enhancements) and the higher luminosity of 
the J-stars (typical of the AGB phase). The large Li abundances typically found in J-type stars ($\sim 80$\% show log $\epsilon$(Li)$>0.5-1.0$, 
Paper I) are consistent with the Li excess that we have found here in early-R stars, supporting this idea. 
Nevertheless, this picture is far from clear since, for instance, the galactic distribution of the J-type stars 
is different from that of the early-R stars: J-type stars are located mainly in the galactic 
thin disk (only $\sim 17\%$ of the J-type stars studied by \citealt{chen2007} have galactic latitude values 
$\mid$b$\mid$ $\geq$ 25$\degr$, typical of the thick disk stars). Secondly, a few J-type stars 
have been found in binary systems (e.g. BM Gem, EU And, UV Aur and UKS-Cel, 
\citealt{barnbaum1991,belczynski2000}). Further work is necessary to determine if early-R stars are, in fact, 
the progenitors of some of the J-type stars.

The solution seems to be related with the third scenario during or after the He-flash. However, as
commented in Sect. \ref{intro}, one-dimensional (1D) attempts and very recent three-dimensional (3D) numerical simulations of the He-flash for stars with a metallicity close to solar, as observed in early-R stars 
\citep{dearborn2006,lattanzio2006,mocak2008a,mocak2008b}, have fail to mix carbon with the envelope. Binary mergers have been suggested as the mechanism able to provoke an anomalous He-flash. In the merger, a fast rotating He core is supposed to be formed with, as a consequence, a strong off-center He-flash. This stronger He-flash eventually would
mix carbon to the surface. \citet{izzard2007} explored statistically the different binary scenarios
that might produce a R star. They discussed actually eleven possible channels, the most favourable cases being 
the merger of a He white dwarf with a RGB star and that of a RGB star with a Hertzsprung gap star. In fact, their
binary population study  produces ten times as many stars as required to match the early-R to red clump ratio
observed. This discrepancy (that becomes more severe considering our results) was, however, interpreted by these 
authors as a positive result since they expect that only a very small fraction of the favourable systems would
ignite He while rotating rapidly enough to provoke the conditions for carbon mixing into their envelope. We have tested this scenario by parametrised 1D simulations of the mer\-ging and, for the first dynamical part, by 3D
Smoothed Particle Hydrodynamics (SPH) simulations. An extended discussion of these numerical simulations will be presented in an accompanying paper \citep{piersanti2009}, but we can advance here that in the most favourable merging channels according to \citet{izzard2007}, we do not obtain any carbon mixing \citep[see also][]{zamora2009}. 

%Our 1D numerical simulations lead us to conclude that these mer\-gers do not favour a strong flash. 
%The accreted matter changes the temperature structure of the He-core, that is also less dense due to rotation. For
%initially small He cores ($\sim$0.2 M$_{\sun}$), He is ignited very close to the center and in 
%less degene\-rate conditions as in standard evolution. For
%initially more massive He-cores (0.4--0.5 M$_{\sun}$), He-ignition occurs
%in the recently accreted layers. However in this case, the He-flash is 
%very weak and no mixing is induced. Finally, we were able to perform 3D SPH simulations 
%for the first hours of the merging. These simula\-tions allow us to check the validity of our hypothesis in  
%the 1D simulations: rapid rigid-rotating He-core (nearly at break-up velocity), 
%and a maximum temperature lower than that needed for He-ignition.

Very recently, \citet{wallerstein2009} have found several carbon-rich (C/O $>1$) RR Lyrae stars with similar
N enhancements to those in early-R stars and have suggested that carbon might have mixed to the surface
during the He-flash by an still unknown physical conditions. More theoretical work on this subject is needed 
to determine such conditions.

\section{Summary}

We have shown that the early- and late-type R stars have different properties including
their chemical composition. Late-R stars have almost identical chemical figures
than normal (N-type) AGB carbon stars, they occupy a similar position in the HR diagram, are long
period variables and present infrared excesses due to dust. These stars probably mark the bottom of the
AGB phase and can be considered thus identical to the N stars. For the early-type R stars we confirm the 
chemical features found in the early analysis by \citet{dominy1984} and the fact that many CH-type and/or K giants 
have been erroneously classified as early-type R stars in the available carbon stars catalogues. So early-type R stars 
are indeed rare objects accounting a much smaller fraction among the giant carbon stars than previously
thought. The location of these stars in the red clump provides strong support for their formation through 
an anomalous He-flash that would mix carbon to the surface. However, all the theoretical attempts made up to date have
fail to reproduce such mixing. The origin of early-type R stars still remains a mystery.

\begin{acknowledgements}
      Part of this work was supported by the Spanish Ministerio de Ciencia e Innovaci\'on projects
      AYA2002-04094-C03-03 and AYA2008-04211-C02-02. OZ acknowledges support by the FPI grant and the Plan propio
of University of Granada. We thank the referee,  Dr.  Jorissen, whose detailed comments have helped us to improve the paper. Based on observations collected at the Centro Astron\'omico Hispano Alem\'an (CAHA) at 
Calar Alto, operated jointly by the Max-Planck Institut f\"ur Astronomie and the Instituto de Astrof\'\i sica de Andaluc\'\i a (CSIC).
This work has made use of the SIMBAD database operated at CDS, Strasbourg, France.

\end{acknowledgements}

\bibliographystyle{aa} %style aa.bst
\bibliography{rstars.bib} %your references yourfile.bib

\end{document}